\documentclass[aps,prd,twocolumn,superscriptaddress,longbibliography,nofootinbib]{revtex4-2}
\usepackage[utf8]{inputenc}
\usepackage{color}
\usepackage{graphicx}
\usepackage{subfigure}
\usepackage{xcolor}
\usepackage[normalem]{ulem}
\usepackage[pdftex,breaklinks,colorlinks,linkcolor=blue,citecolor=teal,anchorcolor=red,urlcolor=cyan]{hyperref}
\usepackage{multirow}
\usepackage{orcidlink}

\usepackage{amsmath,amssymb,amsfonts}

\def\prl{Phys. Rev. Lett.}
\def\prd{Phys. Rev. D}

\def\apj{Astrophys. J.}

\def\apjl{Astrophys. J. Lett.}
\def\aap{Astronomy and Astrophysics}
\def\pr{Phys. Rev.}

\def\pau_p{Prog. Theor. Phys.}

\def\mnras{Mon. Not. R. Astron. Soc.}

\def\apss{Astrophys. Space Sci.}
\def\physrep{Phys. Rep.}
\def\nat{Nature}
\def\sovast{Soviet Ast.}
\def\jcap{J. Cosmology Astropart. Phys}

\newcommand{\ourcomment}[1]{}

\def\eq{Eq.}
\def\eqs{Eqs.}
\def\m{\bar m}

\def\I{\mathcal{I}}

\def\e{{\mathcal E}}

\def\mach{\mathcal M}
\def\N{{\mathcal N}}
\begin{document}

\title{Primordial black holes captured by neutron stars: relativistic point-mass treatment}

\author{Thomas W.~Baumgarte\orcidlink{0000-0002-6316-602X}}
\email{tbaumgar@bowdoin.edu}
\affiliation{Department of Physics and Astronomy, Bowdoin College, Brunswick, Maine 04011, USA}

\author{Stuart L.~Shapiro\orcidlink{0000-0002-3263-7386}}
\email{slshapir@illinois.edu}
\affiliation{Department of Physics, University of Illinois at Urbana-Champaign, Urbana, Illinois 61801}
\affiliation{Department of Astronomy and NCSA, University of Illinois at Urbana-Champaign, Urbana, Illinois 61801}

\begin{abstract}
Primordial black holes (PBHs), if they exist, may collide with and be captured by neutron stars.  We adopt a relativistic point-mass approximation to study this capture, the subsequent confinement of the PBH of mass $m$ inside the neutron star of mass $M_* \gg m$, and the PBH's growth by accretion of stellar material.  Building on earlier treatments we systematically study the capture, confinement, and accretion process, characterize the emitted quasiperiodic continuous gravitational-wave signal, track the evolution of the PBH's orbital parameters, and compare the effects of different choices for the prescription of the dissipative forces.  Our point-mass treatment here is applicable in the limit of small PBH masses, for which its effects on the neutron star can be ignored.  
\end{abstract}

\maketitle

%
\section{Introduction}
%

Primordial black holes (PBHs) are hypothetical objects that may have formed in the early Universe.  First proposed by \cite{ZelN67,Haw71} (see also \cite{CarH74}), they have never been observed directly, but could provide a compelling explanation for the dark matter in the Universe.   While a significant contribution of PBHs to the dark matter has been ruled out in some ranges of PBH masses, they remain plausible in other mass windows (see, e.g., \cite{Khl10,CarKSY21,CarK20,CarCGHK24} for reviews; see also \cite{MonCFVSH19}). No constraints currently exist in the window between about $10^{-16} M_\odot$ and $10^{-10} M_\odot$, meaning that PBHs in this mass range could contribute to or even make up the dark-matter content of the Universe (see, e.g., Fig.~10 in \cite{CarKSY21} and Fig.~1 in \cite{CarK20}).  Following \cite{CarK20} we will refer to this mass window as ``window A".   A second ``window B" that is relevant for our work here exists around $M \simeq 10^{-6} M_{\odot}$, even though PBHs in this mass range are not expected to make up more than about 10~\% of the Universe's dark matter content.

If PBHs exist, some of them are likely to collide with other celestial objects.  While the expected event rates are small (see, e.g., \cite{Abretal09,CapPT13,HorR19,ZouH22,CaiBK24}) they depend strongly on a number of assumptions and may be more favorable in special environments, e.g.~globular clusters and galactic centers.   While we focus on direct collision here, PBHs may also be captured 
by means of other processes (e.g.~\cite{BamSDFV09,PanL14,HorR19,GenST20}).  The process of PBH collision or capture has been invoked as a possible source of a number of different astronomical phenomena.  A collision with the Earth, for example, has been suggested to explain the 1908 Tunguska event in Siberia \cite{JakR73} (but see \cite{BeaT74}), while other interactions have been invoked to trigger neutron star implosions and ``quiet supernovae" \cite{BraL18}, fast radio bursts \cite{AbrBW18,AmaS23}, the formation of low-mass stellar black holes \cite{AbrBUW22,OncMGG22}, and the collapse to supermassive black holes (e.g.~\cite{BamSDFV09}), possibly via the formation of PBH clusters \cite{BelDEGGKKRS14,BelDEEKKKNRS19}.  Gravitational-wave signatures of PBHs have been surveyed recently in \cite{Bagetal23}, and the possibility of using solar-system ephemerides to detect PBHs has been discussed in \cite{BerCDVC23,TraGLK23} and references therein.

If the PBH loses a sufficient amount of energy during the first passage through a star it remains gravitational bound.  We will refer to this as the {\em capture} of the PBH.  It may still reemerge from the star, possibly even many times, but will always return, lose more energy in each subsequent passage, until its energy is so small that it can no longer emerge from what has now become a host star.  We refer to this as the {\em confinement} of the PBH.  As has been pointed out by a number of authors (see, e.g., \cite{CapPT13,AbrBW18,GenST20,AbrBUW22,CaiBK24} and references therein), and as we will review in Section \ref{sec:timescales} below, neutron stars are the most likely candidates to capture and confine a PBH.  Once it is confined, the PBH and its host star co-evolve as a ``Thorne-\.Zytkow-like-Object" (TZlO),\footnote{While Thorne-\.Zytkow objects (see \cite{ThoZ75,ThoZ77}) refer to red giants that host a neutron star, we follow \cite{PasLES11} and refer to TZlOs as a larger class of stars that host a compact parasite:  here a neutron star with a small black hole inside.} as the PBH accretes stellar material, grows, and ultimate triggers the host star's dynamical collapse (see \cite{EasL19,RicBS21b,SchBS21} for numerical simulations for black holes located at the neutron star's center).

This paper is the first of two in which we report on a systematic, relativistic study of the capture and confinement of a PBH by a neutron star, together with the subsequent co-evolution leading up to the neutron star's collapse.  In Section \ref{sec:timescales} below we review several timescales governing these processes, and demonstrate that, for the most promising PBH masses, these timescales form a hierarchy spanning vastly different orders of magnitudes.  Therefore, different computational approaches are required to study different aspects of the above processes.  

In this paper we adopt a relativistic point-mass treatment that is adequate while the PBH is small and its effect on the neutron star can be neglected.  Expanding treatments by other authors (e.g.~\cite{CapPT13,AbrBW18,GenST20,ZouH22,GaoDGZZZ23})  we solve the relativistic geodesic equations -- supplemented with small corrections that account for the secular dissipative forces -- in neutron-star spacetimes.  Adopting this approach we explore the capture and confinement process for different initial impact angles, track the PBH's growth using a self-consistent treatment of the accretion process that is suitable for the stiff equations of state governing neutron stars (see \cite{RicBS21a}), study the unique characteristics of the emitted gravitational wave signal, follow the evolution of the orbit's eccentricity, and examine the effects of different prescriptions for the dissipative forces.

In particular, we find that the dissipative forces do not lead to a significant reduction in the PBH's orbital eccentricity.  This is important, because we have previously shown that the pericenter advance of eccentric orbits, whose rate depends on the neutron star structure, leads to beats in the emitted gravitational wave signal \cite{BauS24b}.  A future measurement of the beat frequencies could therefore provide constraints on the nuclear equation of state.

In \cite{BauS24c} we consider PBHs with larger mass, so that its effects on the neutron star cannot be neglected.  We develop numerical tools required to model the interactions between the two objects self-consistently in general relativity, and perform numerical simulations to explore the onset of dynamical instability triggering the neutron star's collapse to a black hole.  In particular we demonstrate that the initial passage of a sufficiently small PBH through a neutron star leaves the latter stable and does not induce collapse and we confirm, to within a factor of a few, the rough estimates of Section \ref{sec:timescales} below for the onset of collapse.

This paper is organized as follows.  After reviewing key features and timescales in Section \ref{sec:timescales} adopting Newtonian formulae, we describe our relativistic point-mass treatment, including the stellar background model and the geodesic equation, in Section \ref{sec:pointmass}.  We discuss our treatment of dissipative forces in Section \ref{sec:dissipation}, and details of our numerical approach in Section \ref{sec:numerics}.  In Section \ref{sec:results} we present results from our integrations, detailing our findings for the capture and confinement of the PBH, the accretion process, characteristics of the gravitational-wave signal, the evolution of eccentricity, and the effects of choices in the treatment of dissipative forces.  We conclude with a brief discussion in Section \ref{sec:discussion}.  We also include four appendices, namely a description of our treatment of orbits in the stellar exterior (Appendix~\ref{sec:exterior}), our implementation of radiation reaction effects (Appendix~\ref{sec:radreac}), an analytic estimate of the PBH's Mach number in the stellar interior (Appendix~\ref{sec:mach}), as well as a motivation for the power-law (\ref{m_versus_a_fit}) between the PBH's growing mass and its orbit's shrinking semi-major axis (Appendix~\ref{sec:m_versus_a}).

While the astrophysical motivation for our discussion here stems from exploring the interaction of PBHs with neutron stars, we note that most of our formalism similarly applies to other TZlOs, i.e.~small ``parasitic" compact objects (possibly a black hole, be it primordial in nature or not, or a neutron star or white dwarf) embedded in a larger stellar ``host" object (possibly a white dwarf, or a main-sequence, massive or supermassive 
star rather than the neutron star considered here). 

%
\section{Timescales and estimates}
\label{sec:timescales}
%

In this Section we briefly review some general features related to the capture of small black holes by larger host stars, similar to the treatment by several other authors (see, e.g., \cite{CapPT13,AbrBW18,GenST20,RicBS21b,AbrBUW22} and references therein).   While we will adopt a relativistic treatment in the following Sections, here we use Newtonian expressions for simplicity.


We denote to the relative speed between the PBH and host star, when they are at large separation, as $v_{\infty}$.  We will discuss possible values for $v_\infty$ below; for now we will only assume that the black hole's kinetic energy
\begin{equation} \label{K_infty}
K_\infty = \frac{1}{2} m v_\infty^2
\end{equation}
associated with $v_\infty$ is much less than the potential energy at the host stellar surface, 
\begin{equation} \label{U_surf}
U_{\rm surf} = - \frac{G m M_*}{R_*},
\end{equation}
in which case the initial impact speed $v_0$ is approximately the host star's escape speed
\begin{equation} \label{escape_speed}
v_0 \simeq v_{\rm esc} = \left( \frac{2 G M_*}{R_*} \right)^{1/2}.
\end{equation}
Here $M_*$ and $R_*$ are the host's mass and radius, and we have assumed that the PBH's mass $m$ and radius are much smaller than those of the host.  Note also that we have neglected any energy loss during the approach, which is appropriate for a direct collision (rather than, e.g., a binary inspiral driven by gravitational-wave emission).

The timescale $\tau_{\rm trans}$ for the PBH's initial transit through the host star is then approximately
\begin{equation} \label{transit_time}
\tau_{\rm trans} \simeq \frac{R_*}{v_{\rm esc}} \simeq \left(\frac{R_*^3}{GM_*} \right)^{1/2} \simeq \tau_{\rm dyn},
\end{equation}
and hence similar to the host's dynamical timescale $\tau_{\rm dyn}$ as it's motion is barely impeded on its initial passage (we neglect several factors of order unity in \eq~\ref{transit_time}).

For simplicity we assume that the host is governed by a polytropic equation of state (EOS)
\begin{equation} \label{eos}
P = \kappa \rho_0^{\Gamma},
\end{equation}
where $P$ is the pressure, $\rho_0$ the rest-mass density, $\kappa$ is a (dimensional) constant, and the adiabatic exponent is given by $\Gamma = 1 + 1/n$, where $n$ is the star's polytropic index. While we do not distinguish between $\rho_0$ and the total mass-energy density $\rho$ in the Newtonian estimates of this Section, we introduce the relativistic notation here already for usage in later Sections.  We may then approximate the sound speed $a$ at the stellar center from
\begin{equation} \label{sound}
a_c = \left( \frac{d P_c}{d \rho_c} \right)^{1/2} = \left( \frac{ \Gamma P_c}{\rho_c} \right)^{1/2} \simeq \left( \frac{\Gamma G M_*}{R_*} \right)^{1/2},
\end{equation}
where we have estimated $P_c \simeq \rho_c G M_*/R_*$ from hydrostatic equilibrium in the last relation.   Comparing \eqs~(\ref{escape_speed}) and (\ref{sound}) we see that the black hole will (initially) enter the star with a speed similar to the sound speed at the stellar center.  It is thus likely that it will be moving at supersonic speeds through the entire host, except possibly at the center of a neutron star governed by stiff EOSs with a large value of $\Gamma$ (see the analytical estimates in Appendix \ref{sec:mach}, as well as the discussion and Fig.~1 in \cite{GenST20}). 

Assuming supersonic motion, the drag force acting on the black hole is expected to be dominated by ``hydrodynamical drag" (i.e.~gravitational deflection of gas at large distance from the black hole) rather than ``accretion drag'' and can be estimated from 
\begin{equation} \label{drag}
F_{\rm defl} \simeq \frac{4 \pi \rho G^2 m^2}{v^2}  \ln \Lambda,
\end{equation}
where the Coulomb logarithm $\ln \Lambda$ is sometimes approximated to take a value of about 10 or so (see, e.g., \cite{Cha43,RudS71}; see also Section \ref{sec:dynfric} below for a more thorough discussion, as well as Fig.~\ref{fig:initialforces} for a confirmation of the above assumption). 

From (\ref{drag}) we may estimate the dynamical friction timescale
\begin{equation} \label{tau_deflect}
\tau_{\rm defl} \simeq \frac{m v}{F_{\rm defl}} \simeq \frac{1}{\ln \Lambda} \frac{M_*}{m} \, \tau_{\rm trans},
\end{equation}
where we have approximated the density by its average value 
\begin{equation} \label{average_density}
\rho_{\rm ave} \simeq \frac{3 M_*}{4 \pi R_*^3}
\end{equation}
and used (\ref{escape_speed}).  For $m \ll M_*$ we have $\tau_{\rm defl} \gg \tau_{\rm dyn} \simeq \tau_{\rm trans}$, already indicating that a straight-forward integration of the dynamical equations is too long to be practical.

From the drag force (\ref{drag}) we may estimate the the PBH's loss of energy during its initial passage through the host from
\begin{align} \label{energy_loss}
\Delta E & \simeq 2 R_* F_{\rm defl} \simeq \frac{8 \pi \rho G^2 m^2 R_*}{v_{\rm esc}^2} \ln \Lambda 
\simeq \frac{3 m^2 \ln \Lambda}{2 M_*} v_{\rm esc}^2, 
\end{align}
where we have again used (\ref{escape_speed}) and (\ref{average_density}).  For the PBH to be {\em captured} on the initial passage, this energy loss must be greater than the kinetic energy at large separation (\ref{K_infty}), which results in the condition
\begin{equation} \label{capture}
\frac{m}{M_*} \gtrsim \frac{1}{3 \ln \Lambda} \left( \frac{v_\infty}{v_{\rm esc}} \right)^2. ~~~~~~~~~~\mbox{(capture)}
\end{equation}
We see that capture is more likely for host stars with large escape speeds $v_{\rm esc}$, which makes neutron stars the most promising candidates for capturing small, and possibly primordial, black holes.   For a neutron star with compaction $GM/(c^2R) \simeq 0.2$, say, the escape speed (\ref{escape_speed}) is approximately $1.9 \times 10^5$ km/s.  

There is significant uncertainty in terms of what might be an appropriate value for $v_\infty$.  For example, if the black hole is a dark-matter particle associated with the dark-matter halo around our Galaxy, then $v_\infty$ can be estimated from the local rotational speed of the Galaxy, e.g.~about 230 km/s in the solar neighborhood.   Alternatively, one might estimate $v_\infty$ from a typical Galactic peculiar speed, $v_\infty \simeq 10$ km/s, again in the solar neighborhood. Adopting this latter value together with $M_* \simeq M_{\odot}$, the condition (\ref{capture}) yields $m \gtrsim 10^{-10} M_{\odot}$.  Evidently, even neutron stars can capture PBHs in the mass ``window A" only if their asymptotic speed $v_\infty$ is small compared to the estimates above.

Once captured, i.e.~gravitationally bound, the PBH may still reemerge from the host, but does not have enough energy to escape to infinity and may hence reenter the host.  Once the black hole is completely absorbed by the star, so that it no longer re-emerges, its energy is similar to the potential energy at the stellar surface (\ref{U_surf}).  We can therefore estimate the number $\N$ of passages before the black hole is completely confined from
\begin{equation} \label{N_estimate}
    \N \simeq \frac{|U_{\rm surf}|}{\Delta E} \simeq \frac{1}{\ln \Lambda} \frac{M_*}{m}.
\end{equation}
Since each orbit takes a time $\tau_{\rm orbit}$ that, 
prior to confinement, is much greater than the transit time $\tau_{\rm trans} \simeq \tau_{\rm dyn}$, we see that the time to trap the black hole inside the star is
\begin{equation} \label{tau_absorb}
\tau_{\rm confine} = \N \tau_{\rm orbit} \gg \N \tau_{\rm dyn} \simeq \frac{1}{\ln \Lambda} \frac{M_*}{m} \tau_{\rm dyn}.
\end{equation}
We again see that, for $m \ll M_*$ it will take {\em many} dynamical timescales until the black holes is completely trapped by the star, making a direct integration impractical.

A condition for the black hole to be confined by the host even upon the first transit results from setting $\N = 1$ (or less!) in (\ref{N_estimate}), so that
\begin{equation} \label{confinement}
\frac{m}{M_*} \gtrsim \frac{1}{3 \ln \Lambda} ~~~~~~~\mbox{(confinement after first transit).}
\end{equation}
Evidently, (\ref{confinement}) is a significantly more stringent condition than (\ref{capture}), and, in fact, usually violates condition (\ref{dynamical_stability}) below.  We therefore conclude that it is highly unlikely that a captured black hole could be completely confined by the host star after the first transit.   Instead, it is much more likely that it emerges from the star but then returns, loses more energy during subsequent transits, and is confined to the host only after approximately $\N$ transits (see also \cite{AbrBW18} for an illustration).  

Once the now {\em endoparasitic} PBH is confined to the host star, and assuming that the host star remains dynamically stable, the two objects coexist as a ``Thorne-Zytkow-like-Object".   The evolution of this system is then driven by the black hole's accretion of stellar material from the host in addition to the dynamical deflection drag.  While the black hole is moving supersonically, the accretion can be described by Bondi-Hoyle-Lyttleton accretion (\cite{HoyL39,BonH44}; see also \cite{Edg04} for a review as well as \cite{PetSST89,BloR12,KaaMCLT22} and references therein for numerical exploration).   As the frictional forces slow down the PBH, however, its motion ultimately becomes subsonic, in which case the accretion process becomes increasingly spherical near and within the accretion radius, hence is adequately described by Bondi accretion (\cite{Bon52,Mic72}, see also \cite{ShaT83} for a textbook relativistic treatment for $1 \leq \Gamma \leq 5/3$). However, evaluating the steady-state relativistic Bondi accretion rate for stiff EOSs with $\Gamma > 5/3$ requires a special treatment and results in
\begin{equation} \label{bondi}
\dot m_{\rm sph} = \frac{4 \pi \lambda_{\rm GR} \rho_0 G^2 m^2}{a^3}.
\end{equation}
As we discuss in more detail in Section \ref{sec:accretion} below, the dimensionless ``accretion eigenvalue" $\lambda_{\rm GR}$ depends on both the EOS and the asymptotic sound speed $a$, and, for $\Gamma > 5/3$, approaches zero as $a \rightarrow 0$ (see also \cite{Beg78,ChaMS16,RicBS21a}). For example, for $\Gamma = 2$, $\lambda_{\rm GR} = 1.49 \, a$ for $a \rightarrow 0$, whereby $\dot m_{\rm sph} \propto \rho_0 m^2/a^2$.  From (\ref{bondi}) we may now estimate the accretion timescale $\tau_{\rm acc}$
\begin{equation} \label{tau_accretion}
\tau_{\rm acc} \equiv \frac{m}{\dot m} = \frac{a^3}{4 \pi \lambda_{\rm GR} G^2 m \rho} 
\simeq \frac{\Gamma^{3/2}}{3 \lambda_{\rm GR} \delta} \frac{M_*}{m} \tau_{\rm dyn},
\end{equation}
where we have adopted central values for both the sound speed $a$, given by \eq~(\ref{sound}), and the density $\rho$, which introduces the 
central condensation parameter $\delta \equiv \rho_c / \rho_{\rm ave}$.  For typical values at the center of neutron stars, $\lambda_{\rm GR}$ is of order unity; see Fig.~3 in \cite{RicBS21a}.  Clearly, steady-state accretion is only possible as long as the accretion timescale $\tau_{\rm acc}$ is longer than the host star's dynamical timescale $\tau_{\rm dyn}$, resulting in the condition
\begin{equation} \label{dynamical_stability}
\frac{m}{M_*} < \frac{\Gamma^{3/2}}{3 \lambda_{\rm GR} \delta} ~~~~~~~(\mbox{dynamical stability}).
\end{equation}
Once the black hole's mass $m$ exceeds this limit, it will swallow the remain host in ``one last gulp"; this criterion therefore determines the endpoint of the TZlO's evolution.   Similarly, a PBH violating this condition even initially will swallow the entire host star before completing a first transit through the star.  For typical neutron star values, the right-hand side of condition (\ref{dynamical_stability}) yields $\sim 0.1$, but numerical simulations of this process (see \cite{BauS24c}) suggest that this is a slight over-estimate and that the dynamical accretion of the host star occurs for somewhat smaller mass ratios, justifying our conclusion above that conditions (\ref{confinement}) and (\ref{dynamical_stability}) are likely to be mutually exclusively.

The estimates of this Section demonstrate that the modeling of the capture and confinement of a small black hole by a neutron star, followed by the consumption of the host star by the now endoparasitic black hole, involves a hierarchy of vastly differing timescales
\begin{equation} \label{tau_hierarchy}
\tau_{\rm confine} \gg \tau_{\rm acc} \gg \tau_{\rm trans} \simeq \tau_{\rm dyn} \gg \tau_{\rm BH},
\end{equation}
where $\tau_{\rm BH} \equiv G m / c^3$ is the timescale associated with the black hole.  Clearly, the range of these scales poses a formidable problem, and makes it impossible to model every aspect of the problem with the same approach.  

Extending work by other authors (e.g.~\cite{CapPT13,AbrBW18,GenST20}) we adopt in this paper a point-mass approximation to model both the orbital dynamics and ``secular" aspects of the problem, including the initial capture of the black hole, its absorption by the star, and its subsequent growth via accretion.  Unlike most previous treatments, however, we adopt a general relativistic framework for this calculation and model the black hole as closely following a geodesic in the spacetime created by a relativistic host, with dynamical friction, accretion and gravitational radiation reaction resulting in small deviation from exact geodesic motion.  Our results confirm and refine the estimates of this Section, but also predict a number of characteristics of the process, for example the evolution of the eccentricity during the black hole's spiral towards the stellar center, the precession of the orbit and properties of the emitted gravitational wave signal.

In \cite{BauS24c} we adopt self-consistent numerical-relativity simulations in order to explore dynamical aspects of the problem.  In particular, we simulate initial transits of the black hole through the star in order to demonstrate that the star can indeed remain stable, and model the final inspiral phase of the black hole together with the dynamical response and final consumption of the host star.

%
\section{Relativistic point-mass treatment}
\label{sec:pointmass}
%

For the remainder of this paper we treat the PBH as a point mass $m$ and ignore its effects on the host star, which we model as a static equilibrium configuration, and whose mass $M_*$ we assume to be much greater than $m$.  Throughout the rest of the paper we adopt geometrized units with $G = 1 = c$ unless noted otherwise.

%
\subsection{Stellar background model}
\label{sec:background}
%

We construct a stellar background model by solving the Oppenheimer-Volkoff (OV) equations (see \cite{OppV39}) for a star governed by a polytropic EOS (\ref{eos}).    Specifically, the enclosed gravitational mass $M(R)$ satisfies
\begin{equation} \label{TOV_1}
    \frac{dM}{dR} = 4 \pi \rho R^2 dR,
\end{equation}  
while pressure $P$ satisfies
\begin{equation} \label{TOV_2}
\frac{d P}{dR} = - (\rho + P)\frac{M(R) + 4 \pi P R^3}{R^2 - 2 M(R) R},
\end{equation}
where $R$ is the areal radius.  We integrate the above equations from $R = 0$ to the point at which $P$ vanishes, which defines the stellar radius $R_*$ and its total gravitational mass $M_* = M(R_*)$.  Assuming that the star is isentropic, the total energy density $\rho$ is related to the rest-mass energy density $\rho_0$ by $\rho = \rho_0 + P/(\Gamma - 1)$ in the above equations.

We next transform the solution to an isotropic coordinate radius $r$, which brings the metric into the form
\begin{equation} \label{TOV_metric_isotropic}
ds^2 = - \alpha(r)^2 dt^2 + A(r)^2 (dr^2 + r^2 d\Omega^2).
\end{equation}
Here $\alpha$ is often referred to as the lapse function, and $A$ is related to the conformal factor, which is more commonly written as $\psi = A^{1/2}$.  In the stellar exterior, these metric coefficients take the form
\begin{subequations} \label{metric_coefficients_exterior}
\begin{align}
\alpha(r) & = \frac{1 - M/(2r)}{1 + M/(2r)} \\
A(r) & = \left(1 + \frac{M}{2r}\right)^2,
\end{align}
\end{subequations}
while in the interior they are computed from the OV solution and matched to the exterior solution 
(\ref{metric_coefficients_exterior}) at the stellar surface $r_*$ (see, e.g., \cite{BauS24b} for details).

The radiation-reaction forces (\ref{AFRR1}) also require the first three derivatives of the enclosed mass $M(r)$ with respect to $r$.  In the stellar exterior these derivatives all vanish, and in the interior we compute them using Eqs.~(\ref{TOV_1}) and (\ref{TOV_2}).

In this paper we focus on $\Gamma = 2$ polytropes as plausible models of neutron stars with a moderately stiff EOS.  We vary the central density $\rho_c$ so that star's compaction takes the value $M_*/R_* = 1/6$, which is consistent with, for example, NICER observations of the millisecond pulsar PSR J0030+0451 \cite{NICER_J0030}. 

%
\subsection{Equations of motion}
\label{sec:eom}
%

In the absence of any retarding forces the PBH can be treated like a test particle freely falling in the geometry of the spacetime (\ref{TOV_metric_isotropic}), in which case its orbit follows a geodesic.   In reality, the PBH experiences dissipating forces resulting from dynamical friction, accretion drag, and radiation reaction.  Since all three are very small and affect the orbit only on time scales significantly longer than the orbital timescale, at least for sufficiently small black holes, we treat these retarding forces as small perturbations and add them as small corrections to the right-hand sides of the geodesic equations. 

Since the spacetimes (\ref{TOV_metric_isotropic}) possess both a time-like Killing vector $\xi_{(t)} = \partial/\partial t$ and a space-like Killing vector $\xi_{(\varphi)} = \partial/\partial \phi$, we can define the PBH's energy per unit mass as
\begin{equation} \label{define_e}
\e \equiv - u_a \xi^a_{(t)} = - u_t 
=  \alpha^2 \frac{dt}{d\tau} 
= \mbox{const},
\end{equation}
and the angular momentum per unit mass as
\begin{equation}\label{define_l}
\ell \equiv u_a \xi^a_{(\varphi)} = u_\varphi 
= r^2 \sin^2 \theta A^2 \frac{d\varphi}{d\tau} 
= \mbox{const},
\end{equation}
where $u^a$ is the PBH's four-velocity.  In the absence of dissipative forces $\e$ and $\ell$ are strictly conserved constants of motion.

We choose the equatorial plane of our coordinate system to be aligned with the plane of the orbit, in which case $\theta = \pi/2$ and $u^\theta = 0$.  We then write out the geodesic equations following \cite{ShaT85} and add the dissipative forces on the right-hand side to obtain
\begin{subequations} \label{eom}
\begin{align}
\frac{d r}{d t} = & \, \frac{1}{A^2 u^t} u_r \\
\frac{du_r}{d t} = & \, - \alpha u^t \partial_{r} \alpha 
+ \frac{u_r^2}{u^t}\frac{\partial_{r} A}{A^3} \nonumber \\ 
& \, + \frac{u_\varphi^2}{u^t} \left(\frac{1}{r^3 A^2} + \frac{\partial_{r} A}{r^2 A^3} \right)
+ \frac{\eta_{\rm su}}{m} \left(\frac{d p}{d t} \right)_r^{\rm diss}\label{durdt}  \\
\frac{d\varphi}{d t} = &\, \frac{1}{r^2 A^2 u^t} u_\varphi \\
\frac{d u_\varphi}{d t} = & \,\frac{\eta_{\rm su}}{m} \left(\frac{d p}{d t} \right)^{\rm diss}_{\varphi} \label{duphidt} \\
\frac{dm}{dt} = &\, \dot m.
\end{align}
Here the time component of the four-velocity $u^t = dt/d\tau$ can be computed from the normalization of the four-velocity $u_a u^a = -1$, which yields
\begin{equation} \label{ut}
\alpha u^t = \left( 1 + \frac{u_r^2}{A^2} + \frac{u_{\varphi}^2}{r^2 A^2} \right)^{1/2},
\end{equation}
\end{subequations}
and we provide an expression for the accretion rate $\dot m$ in Section \ref{sec:accretion} below (see Eq.~\ref{mdot_general}).

As we discuss in detail in Section \ref{sec:dissipation} below, we take into account three different contributions to the dissipative forces,
\begin{equation} \label{F_diss}
\left(\frac{d p}{d t} \right)_i^{\rm diss} = 
\left(\frac{d p}{d t} \right)_i^{\rm defl} + 
\left(\frac{d p}{d t} \right)_i^{\rm acc} + 
\left(\frac{d p}{d t} \right)_i^{\rm RR}.
\end{equation}
Here the first term on the right-hand sides accounts for forces resulting from dynamical friction (see Section \ref{sec:dynfric}), the second for accretion forces (Section \ref{sec:accretion}), and the last for gravitational radiation reaction (Section \ref{sec:radiation_reaction} and Appendix \ref{sec:radreac}).

Finally we note that we introduced in (\ref{eom}) a dimensionless {\em speed-up} factor $\eta_{\rm su}$ that allows us to artificially speed up the secular dissipative effects by choosing $\eta_{\rm su} > 1$ (see Section \ref{sec:speedup} below for details).

%
\section{Dissipative forces}
\label{sec:dissipation}
%

In this Section we discuss the three sources of dissipation that we take into account in our simulations: dynamical friction (Section \ref{sec:dynfric}), accretion (Section \ref{sec:accretion}), and gravitational radiation reaction (Section \ref{sec:radiation_reaction}).

%
\subsection{Dynamical friction}
\label{sec:dynfric}
%

Different authors have suggested different approximations for modeling the dissipative effects of dynamical friction (see, e.g., \cite{Cha43,RudS71}).  Here we follow the prescriptions of \cite{PetSST89,Bar07}, both of whom present relativistic expressions for the dynamical friction.  The treatment of \cite{Bar07} generalizes the Newtonian approach of \cite{Ost99,KimK07} for relativistic motion, and also demonstrated agreement with the earlier relativistic results of \cite{PetSST89} for steady-state, straight-line motion and speeds much greater than the sound speed.

Specifically, we start by writing the magnitude of the deflection force as
\begin{equation} \label{F_defl_mag}
\left( \frac{d p}{d \tau_{\rm LAO}} \right)_{\rm defl} =  \frac{4 \pi (\rho + P) m^2 (1 + 2 u^2)^2}{u^2} I
\end{equation}
where $\tau_{\rm LAO}$ is the proper time of a {\em local asymptotic observer} (LAO) and $I$ is a dimensionless function, both of which we will describe below, and where we define
\begin{equation}
u \equiv (\gamma^{ij} u_i u_j)^{1/2} = A^{-1} ((u_r)^2 + r^{-2} (u_\varphi)^2)^{1/2}
\end{equation}
({\em cf.}~Eq.~B.45 in \cite{PetSST89}, who set
$I = \ln {\Lambda}$ where $\Lambda = b_{\rm max}/b_{\rm cut}$ with $b_{\rm max} \sim R$ and $b_{\rm cut} \sim r_a \sim m/v^2$). Using $u^2 = \gamma^2 v^2$ for Minkowski spacetimes, the above expression is equivalent to (66) in \cite{Bar07} for straight-line motion. 

The deflection force is usually derived for a perturber traveling through a homogeneous fluid whose unperturbed density $\rho$ and pressure $P$ are constant in the frame of a static observer at a large distance from the perturber, and whose self-gravity can be ignored in comparison to $m$.  In this case, the time derivative is taken with respect to the proper time of this distant observer, which is simply the coordinate time.

In our application, however, the perturber travels through the interior of a neutron star, so that the density and pressure cannot be considered constant at large distances.  We can apply (\ref{F_defl_mag}) regardless, assuming that the perturber -- the PBH -- is sufficiently small so that we can introduce a hierarchy of length scales $r_{\rm BH} \ll d \ll r_*$.  We then consider a static LAO  whose distance $d$ from the perturber is much greater than the PBH's  horizon radius $r_{\rm BH}$, but also much smaller than the neutron star radius $r_*$.  The (unperturbed) density and pressure can then be approximated as constant in the region between the black hole and the LAO, and we may interpret the deflection force (\ref{F_defl_mag}) as measured by such a LAO.\footnote{See also \cite{RicBS21b}, in particular Appendix A, where we introduced the notion of a LAO in the context of Bondi accretion inside a neutron star and labeled $\tau_{\rm LAO}$ by $\tau_*$.}

In the coordinates of the metric (\ref{TOV_metric_isotropic}), the four-velocity of the LAO is given by $u^a_{\rm LAO} = (\alpha^{-1},0,0,0)$.  The LAO's proper time $\tau_{\rm LAO}$ therefore advances at a rate 
\begin{equation} \label{dtauLAO}
d\tau_{\rm LAO} = \alpha dt
\end{equation}
relative to that of the asymptotic observer at infinity (whose proper time is equal to the coordinate time $t$).   Moreover, in the LAO's orthonormal frame, the components of the PBH's coordinate velocity take the values
\begin{equation} \label{vhats}
v^{\hat r} = \frac{u^{\hat r}}{u^{\hat t}} = \frac{(u_r/A)}{\alpha u^t},~~~~~
v^{\hat \varphi} = \frac{u^{\hat \varphi}}{u^{\hat t}} = \frac{(u_\varphi/r A)} {\alpha u^t}.
\end{equation}
Accordingly, the LAO measures the PBH's speed to be
\begin{equation}
v_{\rm LAO} = ((v^{\hat r})^2 + (v^{\hat \varphi})^2)^{1/2} 
= \frac{u}{\alpha u^t} = \frac{u}{\gamma},
\end{equation}
where we have defined the Lorentz factor between the PBH and the LAO as $\gamma \equiv 
- u_a u^a_{\rm LAO} 
= \alpha u^t$.  We note that we may rewrite (\ref{ut}) in terms of $v_{\rm LAO}$ to obtain the familiar relation $\gamma = \alpha u^t = (1 - v_{\rm LAO}^2)^{-1/2}$.

Using relation (\ref{dtauLAO}) and decomposing (\ref{F_defl_mag}) so that the force points in the direction opposite to the PBH velocity we then obtain its orthonormal components:
\begin{equation} \label{F_defl_ortho}
\left( \frac{d p}{d t} \right)^{\hat \imath}_{\rm defl} = - \frac{4 \pi \alpha (\rho + P) m^2 (1 + 2 u^2)^2}{u^2} I \frac{v^{\hat \imath}}{v_{\rm LAO}}.
\end{equation}
From the orthonormal components (\ref{F_defl_ortho}) we finally compute the covariant coordinate components,
\begin{equation} \label{F_defl}
\left( \frac{d p}{d t} \right)_{i}^{\rm defl} = - \frac{4 \pi \alpha (\rho + P) m^2 (1 + 2 u^2)^2}{u^3} I u_i.
\end{equation}
which, with a suitable choice for the function $I$, we adopt in our simulations.

As our first choice for the function $I$, which we will refer to as our {\em canonical} case, we adopt the expressions of \cite{Ost99},
\begin{equation} \label{dyn_friction_ostriker}
I = \left\{ \begin{array}{ll}
\displaystyle \frac{1}{2} \ln \left( \frac{1+\mach}{1 - \mach} \right) - \mach ~~~ & \mach < 1 \\
\displaystyle \frac{1}{2} \ln \left( 1 - \mach^{-2} \right) + \ln \Lambda 	& \mach > 1.
\end{array}
\right.
\end{equation}
Here $\mach = v_{\rm LAO} / a$ is the Mach number, and $a$ the relativistic sound speed
\begin{equation} \label{a_of_rho}
a^2  = \frac{\Gamma K \rho_0^{\Gamma - 1}}{1 + \Gamma K \rho_0^{\Gamma - 1}/(\Gamma - 1) }.
\end{equation}
Note that (\ref{dyn_friction_ostriker}) reduces to $\ln \Lambda$ in the limit $\mach \gg 1$.  For simplicity we follow other authors (e.g.~\cite{Abretal09}) and approximate $\ln \Lambda = 10$ in the above expression unless noted otherwise.

We note that the limit $\mach \rightarrow 1$ is irregular in  (\ref{dyn_friction_ostriker}): the subsonic expression diverges to positive infinity, while the supersonic expression approaches negative infinity.  In order to avoid this artifact we linearly interpolate between two values of $\mach$ for which the subsonic and supersonic expressions take similar values for $I$. 

As an alternative we adopt the prescription of \cite{PetSST89}.  Specifically, we assume that the drag vanishes for subsonic flows (see also \cite{RepS80}), and we follow \cite{PetSST89} to estimate $\ln \Lambda$ for supersonic flow during the simulation. Specifically, we write
\begin{equation} \label{dyn_friction_petrich}
I = \left\{ \begin{array}{ll}
\displaystyle 0 & \mach < 1 \\
\ln \Lambda 	& \mach > 1,
\end{array}
\right.
\end{equation}
where $\Lambda = b_{\rm max}/b_{\rm min}$ is the ratio between the maximum and minimum impact parameters, and estimate
\begin{equation} \label{Petrich_Lambda}
\Lambda \simeq \frac{R v^2}{m}
\end{equation}
(see \cite{PetSST89} for details).   

As an aside we note that the above expressions assume straight-line motion, and that several authors have also provided prescriptions of dynamical friction for circular orbits (see, e.g., \cite{KimK07,Bar07}).  Here we adopt the straight-line rather than circular expressions, because we find that generic orbits retain considerable eccentricity.  The difference between the two prescriptions essentially amount to the appearance of an extra Lorentz factor in the circular case (see \cite{Bar07}).

%
\subsection{Accretion}
\label{sec:accretion}
%

Following \cite{PetSST89} again, we write the accretion drag force as
\begin{equation} \label{F_acc_1}
\left( \frac{dp}{d\tau_{\rm LAO}} \right)^{\rm acc}_i = - \frac{\rho + P}{\rho_0} \dot m u_i,
\end{equation}
({\em cf.}~Eq.~(2.40) in \cite{PetSST89}).  Here $\dot m$ is the (rest-mass) accretion rate, and, as in Section \ref{sec:dynfric}, we interpret all time derivatives as taken with respect to a LAO. 

Spherically symmetric, adiabatic, steady-state accretion onto a black hole at rest in a medium that is asymtotically homogeneous, and under the assumption that the fluid's self-gravity may be ignored, is described by Bondi accretion (see \cite{Bon52} for a Newtonian treatment, \cite{Mic72} for a relativistic generalization, and \cite{ShaT83} for a textbook treatment of both that corrects earlier errors for $\Gamma = 5/3$ and proves that only the transonic solution is viable for black holes).  The accretion rate is given by \eq~(\ref{bondi}), where $a$ is the asymptotic sound speed at a distance much greater than the accretion radius $r_s = G m / a^2$.  Under the assumption that $m \ll M$ we typically have $r_s \ll R$ and may approximate $a$ as the sound speed in the neighborhood of the black hole's current location $r$, as observed by a LAO.     

For soft EOSs, with $1 \leq \Gamma \leq 5/3$, and in the Newtonian limit, the accretion eigenvalues $\lambda_{\rm GR}$ reduce to their Newtonian counterparts 
\begin{equation} \label{lambda_s}
\lambda = \frac{1}{4} \left(\frac{2}{5 - 3 \Gamma} \right)^{(5 - 3 \Gamma)/(2 \Gamma - 2)} ~~~~~~~\mbox{($\Gamma \leq 5/3$, $a \ll 1$).}
\end{equation}
Evidently this expression breaks down for stiff EOSs with $\Gamma > 5/3$, which instead require a relativistic treatment (see, e.g., \cite{Beg78,ChaMS16,RicBS21a} and references therein).  Within such a relativistic treatment, the values $\lambda_{\rm GR}$ are no longer independent of $a$, and have to be found by solving a cubic equation (see \cite{RicBS21a} for details and their Fig.~2 for examples).  In particular, $\lambda_{\rm GR} \rightarrow 0$ as $a \rightarrow 0$ for $\Gamma > 5/3$.  For $\Gamma = 2$, for example,  one finds
\begin{equation} \label{lambda_GR}
\lambda_{\rm GR} \simeq 1.49 \, a, ~~~~~~~\mbox{($\Gamma = 2$, $a 
\lesssim 0.2$)}.
\end{equation}
Inserting (\ref{lambda_GR}) into (\ref{bondi}) demonstrates that, in the limit $a \ll 1$, the accretion rate becomes
\begin{equation} \label{bondi_Gamma_2}
\dot m_{\rm sph} \simeq 1.49 \frac{4 \pi \rho_0 m^2}{a^2} ~~~~~~~\mbox{($\Gamma = 2$, $a \lesssim 0.2$)}
\end{equation}
and hence scales with $a^{-2}$ rather than $a^{-3}$ as in the Newtonian expressions.  In the low-density limit $\rho_0 \rightarrow 0$, the sound speed (\ref{a_of_rho}) reduces to $a^2 \simeq \Gamma K \rho_0^{\Gamma - 1}$, which we may use in (\ref{bondi_Gamma_2}) to show that the accretion rate $\dot m$ takes a {\em minimum value}
\begin{equation}
\dot m^{\rm min}_{\rm sph} \simeq 9.36 \frac{m^2}{K}
\end{equation}
(see Eq.~(51) as well as Fig.~4 in \cite{RicBS21a}).  Both the existence of this minimum accretion rate, as well as the difference of the scaling with $a$, represent qualitative differences from the Newtonian results (see also \cite{RicBS21b,SchBS21} for numerical confirmations of these results).  

The above expressions are correct only in the limit that asymptotically the fluid is at rest with respect to the black hole.  Clearly this is not the case early on in our simulations when the black hole is moving through the star.  Following several other authors  (see, e.g., \cite{Bon52,ShiMTS85,PetSST89} and references therein),  we approximately account for this effect by multiplying the spherical accretion rate $\dot m_{\rm sph}$ with a Bondi-Hoyle-Lyttleton-like correction factor 
\begin{equation} \label{mdot_general}
\dot m = \dot m_{\rm sph} \left( \frac{a^2}{a^2 + v^2_{\rm LAO}} \right)^{q/2}.
\end{equation}
Motivated by the scaling with $a^{-3}$ of the Newtonian spherical Bondi accretion rate, several previous authors adopted $q = 3$, as argued in the original treatments in 
\cite{BonH44,HoyL39}.  Given the scaling of (\ref{bondi_Gamma_2}) with $a^{-2}$, however, it appears more natural to adopt $q = 2$ for $\Gamma = 2$.
Accordingly, we have 
\begin{equation}
\dot m \simeq 1.49 \frac{4 \pi \rho_0 m^2}{(a^2 + v^2_{\rm LAO})} ~~~~~~~\mbox{($\Gamma = 2$, $a \ll 1$)}
\end{equation}
in the limit $a \rightarrow 0$.  As a canonical model in our simulations we therefore adopt (\ref{mdot_general}) with $q = 2$.

Assembling the above in the accretion drag force (\ref{F_acc_1}), and expressing the time derivative in terms of $t$, we obtain
\begin{equation} \label{f_accretion_final}
\left( \frac{dp}{dt}  \right)^{\rm acc}_i = - \frac{4 \pi \alpha \lambda_{\rm GR} (\rho + P) m^2}{a^3} \left( \frac{a^2}{a^2 + v^2_{\rm LAO}} \right)^{q/2} u_i.
\end{equation}
Here we evaluate $\lambda_{\rm GR}$ self-consistently, as described in \cite{RicBS21a}, adopting $q = 2$ in our canonical model.

Finally we note that the Bondi formula (\ref{bondi}) provides the rate at which the black hole accretes rest mass only, and does not account for any other forms of energy (e.g.~the fluid's internal or thermal energy). The black hole's gravitational mass, which accounts for all forms of energy, therefore grows at a rate that may be greater than that given by the Bondi accretion rate (see \cite{RicBS21b} for more details and numerical examples).  For our purposes here we assume that the difference between the two rates is small and and can be neglected.

%
\subsection{Radiation Reaction}
\label{sec:radiation_reaction}
%

As a third dissipative force we account for the effects of gravitational-wave radiation reaction.   We apply the weak-field, slow-velocity formalism (see \cite{PetM63,Pet64,MisTW73}) for our treatment of the gravitational waves radiation reaction force and the gravitational waves that are generated.  Our prescription is thus most reliable in the stellar far exterior, where the radiation reaction force plays a dominant role, and only approximate in the neutron star interior, where it is dominated by the other dissipative forces.  Our results for the gravitational waveforms arising from the PBH orbital motion are similarly approximate.

We refer the reader to Appendix \ref{sec:radreac} for a derivation of all relevant expressions.  In particular, the radiation-reaction forces $F^i_{\rm RR}$ are listed in Eqs.~(\ref{AFRR1}), from which we identify the terms
\begin{equation}
\left( \frac{dp}{dt} \right)^{\rm RR}_i = \gamma_{ij} F^j_{\rm RR}
\end{equation}
in Eq.~(\ref{F_diss}).

%
\section{Numerical integration}
\label{sec:numerics}
%

\subsection{Initial data}
\label{sec:numerics:indata}

We start the integration of Eqs.~(\ref{eom}) by placing the PBH with an initial mass $m = m(0)$ at an initial radius $r = r(0)$, which unless noted otherwise, we choose to be the neutron star's surface, $r(0) = r_*$.  Without loss of generality we also choose $\varphi(0) = 0$ initially.  We then choose an asymptotic speed $v_\infty$, and use conservation of the energy (\ref{define_e}) to compute the PBH's speed at the stellar surface.  Specifically, we first evaluate 
\begin{equation}
\e = - u_t = \frac{dt}{d\tau} = \gamma = (1 - v^2_\infty)^{-1/2}
\end{equation}
at infinity.  We next evaluate (\ref{ut}) at the initial radius to compute the quantity
\begin{equation}
u \equiv \left(u_r^2 + \frac{ u_\varphi^2}{r(0)^2}\right)^{1/2} 
= A \left( \frac{\e^2}{\alpha^2} - 1 \right)^{1/2},
\end{equation}
where both $A$ and $\alpha$ are evaluated at $r(0)$.  We further choose an initial impact angle $\psi_0$, from which we compute the initial velocity components
\begin{equation}
u_r(0) = - u \cos \psi_0,~~~~~~u_\varphi(0) = r(0) \, u \sin \psi_0.
\end{equation}
Given these initial conditions we integrate Eqs.~(\ref{eom}) with a {\tt python} implementation of a 4th-order Runge-Kutta method.  Assuming that the total energy is negative after the first passage through the star, indicating that the PBH has been captured, we terminate the integration when either the PBH emerges from the star, or when its mass exceeds a maximum value.

\subsection{Exterior evolution}
\label{sec:numerics:exterior}

Even once captured, a PBH (with sufficiently small mass) will spend vastly more time on (bound) orbits in the stellar exterior than in the stellar interior.  In general, therefore, tracking these exterior orbits via direct integration of the equations of motion (\ref{eom}) would be prohibitively expensive.  Information about the exterior parts of the orbits is needed, however, both to compute the total time required to capture a PBH, and to compute the loss of energy and angular momentum via gravitational-wave emission.  For those cases with $m(0)/M_* \ll 10^{-3}$, we therefore treat this part of the problem in an approximate Newtonian framework, so that the exterior trajectory is given by a segment of a Kepler orbit, and solve the equations analytically.

Specifically, when the PBH emerges from the star, we first compute the relativistic energy $\e = - u_t$ (using \ref{ut}) and angular momentum $\ell = u_\varphi$ per unit mass, from which we estimate corresponding Newtonian values according to 
\begin{subequations}  \label{EL_Newton}
\begin{align}
E_{\rm Newt} & = \frac{1}{2} (\e^2 - 1) \\
L_{\rm Newt} & = \ell.
\end{align}
\end{subequations}
Given these, the semi-major axis $a$ and eccentricity $e$ of the exterior orbit are then given by
\begin{subequations} \label{ae_Newton}
\begin{align}
a & = - \frac{M}{2 E_{\rm Newt}} \\
e & = \left(1 + \frac{2 E_{\rm Newt} L_{\rm Newt}^2}{M^2} \right)^{1/2}.
\end{align}
\end{subequations}
Given the above parameters, the time spent in the exterior can be computed by direct integration and is given by (\ref{deltat4}) in Appendix \ref{sec:exterior}.  

As discussed in Appendix \ref{sec:radreac:exterior}, we also compute the loss of energy $\Delta E_{\rm Newt}$ (see \ref{ADeltaE}) and angular momentum $\Delta L_{\rm Newt}$ (see \ref{ADeltaL}) due to the emission of gravitational radiation.  From the latter, we compute updated values of $E_{\rm Newt}$ and $L_{\rm Newt}$ and then invert (\ref{EL_Newton}) to obtain the new values of $e$ and $\ell$ with which the PBH will re-enter the neutron star.  From the latter we immediately obtain $u_{\varphi} = \ell$, and inserting $\e = - u_t$ together with $u_\varphi$ into (\ref{ut}) yields $u_r$ (where we take the negative root), providing the initial data for the next passage through the neutron star.

\subsection{Artificial speed-up}
\label{sec:speedup}

As we discussed in Section \ref{sec:timescales}, the ratio between the dissipative timescale $\tau_{\rm diss}$ and the dynamical and transit timescales $\tau_{\rm dyn} \simeq \tau_{\rm trans}$ scales approximately with the mass ratio $M / m$ (see Eq.~\ref{tau_deflect}).  Even our approximate point-mass treatment is therefore prohibitively slow, in particular for black holes in the mass window A, with $m \lesssim 10^{-10} M$.  We therefore introduced the speed-up factors $\eta_{\rm su}$ in Eqs.~(\ref{eom}), which allow us to artificially speed up the dissipative evolution.   In practice, we start each integration with the physical value of $\eta_{\rm su} = 1$, so that we can reliably determine whether or not the PBH is captured after the first passage.  Moreover, the first few exterior orbits take significantly longer than subsequent ones, because the PBH's energy is still close to zero, its semi-major axis large, and hence the orbital period long.  These orbits would be missed, and hence the capture time underestimated, if we artificially increased the dissipative effects immediately.   Instead, we turn on the speed-up by choosing $\alpha_{\rm su} > 1$ up to a maximum value of $\alpha_{\rm su}^{\rm max}$ only after a specified number of passages through the star.   After turning on the speed-up we multiply all times and passages with $\alpha_{\rm su}$ in order to obtain a good estimate for the true evolutionary times and number of passages through the star.  As the black hole grows, we reduce $\eta_{\rm su}$ again, always ensuring that $\eta_{\rm su} \leq \max(1, 10^{-4} M/ m)$, so that the artificially ``sped-up" dissipative timescale 
\begin{equation}
\tau_{\rm diss, su} = \frac{\tau_{\rm diss}}{\eta_{\rm su}} \simeq \frac{1}{\eta_{\rm su}} \frac{M}{m} \tau_{\rm dyn} \geq 10^4 \tau_{\rm dyn}. 
\end{equation}
remains several order of magnitude greater than the dynamical timescale (as long as $m < 10^4 M$).  This entire approach is possible due to the small perturbative nature of the dissipative effects on dynamical timescales.

\subsection{Termination}
\label{sec:termination}

We terminate the integration when the PBH's mass exceeds the limit obtained from the stability criterion (\ref{dynamical_stability}).   Once the black hole's mass becomes relatively large and approaches this limit accretion proceeds very quickly, so that the precise value adopted for the right-hand side has very little effect on the overall evolutionary time.  Based on our numerical simulations in \cite{BauS24c} we adopt a mass ratio of $10^{-2}$ as an approximate limit, and hence use
\begin{equation} \label{termination}
m = 10^{-2} M
\end{equation}
as our termination criterion.



%
\section{Results}
\label{sec:results}
%

For all results presented in this Section we adopt a $\Gamma = 2$ polytrope to model the neutron star, as discussed in Sect.~\ref{sec:background}, and choose its central density so that the star's compaction becomes $M_*/R_* = 1/6$.  When reporting results in terms of physical units we assume that the neutron star has a mass of $M_* = 1.4 M_\odot$, for which the above compaction yields a radius of $R_* \simeq 12.5$ km.  Both $M_*$ and $R_*$ are well within the error bars of NICER observations of the millisecond pulsar PSR J0030+0451 \cite{NICER_J0030}.  The dynamical timescale for such a star can be estimated to be
\begin{equation} \label{tau_dyn}
\tau_{\rm dyn} \simeq \left( \frac{R_*^3}{G M_*} \right)^{1/2} \simeq 0.1\,\mbox{ms}.
\end{equation}

We also assume for the results presented here that $v_\infty = 1$ km/s (see Section \ref{sec:numerics:indata}).  This value is smaller than, for example, the Sun's speed relative to the local standard of rest by about an order of magnitude (see, e.g., \cite{SchBD10}), and may therefore represent a typical value at the lower end of stellar peculiar speed distributions.

Unless noted otherwise we show results for our ``canonical" choices discussed in Section \ref{sec:dissipation} above, namely the Ostriker prescription (\ref{dyn_friction_ostriker}) for dynamical friction with $\ln \Lambda = 10$ and relativistic Bondi accretion (\ref{mdot_general}) with $q = 2$.

%
\subsection{An example orbit}
\label{sec:results:example}
%

\begin{figure}
    \includegraphics[width = .45 \textwidth]{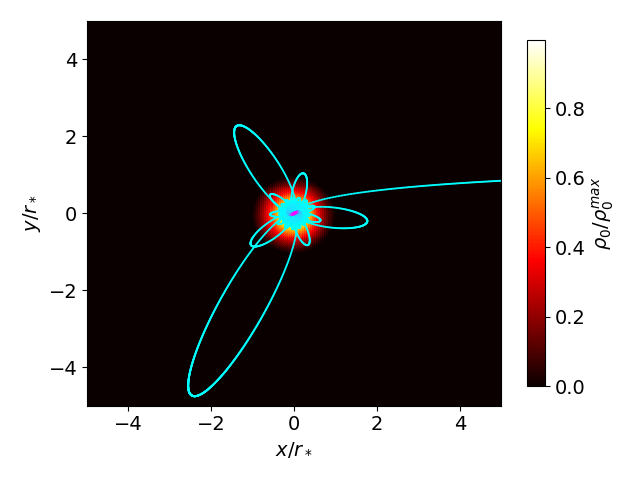}
    \includegraphics[width = .45 \textwidth]{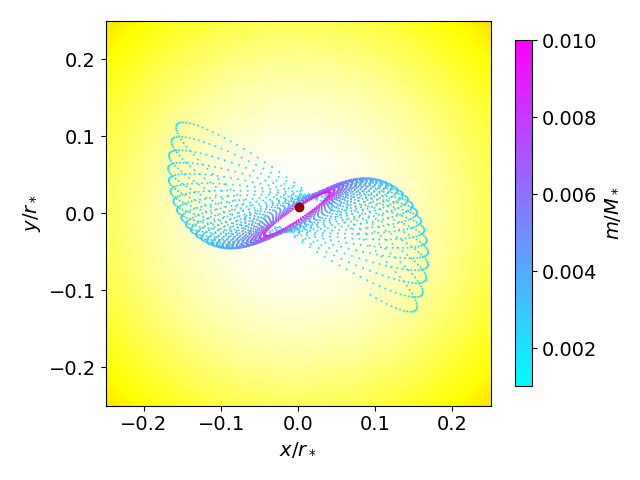}
    \caption{Example capture orbit for an initial PBH mass of $m(0) = 10^{-3} M_*$. The PBH first approaches the star approximately along the positive $x$-direction.  In the upper panel we show the entire orbit, with the background shading indicating the rest-mass density distribution inside the neutron star as indicated by the color bar.  In the lower panel we show the last few orbits only.  The color shading of the PBH trajectory represents its mass, as shown by the color bar in the lower panel, and the individual dots represent increments of time by about 4.6 $\mu$s. We terminate the evolution once $m$ reaches 1\% of the neutron star mass $M_*$, at a location that is indicated by the dot (see also Fig.~\ref{fig:demo_radius_and_mass}).}
    \label{fig:demo_orbit}
\end{figure}

\begin{figure}
    \includegraphics[width = .45 \textwidth]{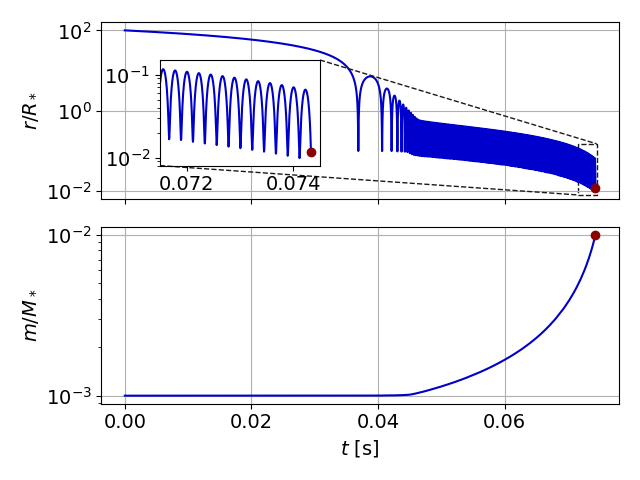}
    \caption{The radius $r$ and mass $m$ for the same example capture orbit as in Fig.~\ref{fig:demo_orbit}.  Early on, while the PBH spends most time outside of the NS, its mass increases only during the brief passages through the star; these small increases are not visible in the plot.  Once completely inside the star, the accretion rate increases rapidly -- faster than exponential -- as $m$ increases.}
    \label{fig:demo_radius_and_mass}
\end{figure}

We start our discussion by showing, for illustrative purposes, an example capture orbit for a PBH with a large initial mass of $m(0) = 10^{-3} M_*$, which allows us to integrate the equations of motion even in the exterior without invoking the estimates of Section \ref{sec:numerics:exterior}, and which results in figures that are not overly crowded.

In Fig.~\ref{fig:demo_orbit} we show the PBH's trajectory.  In both panels the color coding of the background represents the density distribution in the interior of the neutron star.  In the top panel, the PBH arrives from large positive $x$ near the $x$-axis and is deflected towards the bottom left after its first passage through the star.  Now bound gravitationally, the PBH returns to the star and makes several more passages through its interior before remaining completely contained inside the star and no longer emerging from its surface.  In the bottom panel we show the last few orbits before the PBH's mass reaches the termination criterion (\ref{termination}), at a location marked by the dot, and we end the calculation. 

In Fig.~\ref{fig:demo_radius_and_mass} we show the PBH's distance $r$ from the center of the neutron star, as well as its mass $m$, as a function of time.  We start the evolution when the PBH is at a distance of $r = 100 R_*$.  After a time of approximately 0.045 s the PBH remains completely inside the neutron star.  Prior to this time, the mass $m$ increases only by very small amounts (that are not visible on the plot) while the PBH passes through the star; once inside the star, the accretion rate scales with the square of the PBH mass $m$ (see \ref{bondi}), resulting in a faster-than-exponential growth in $m$.  As discussed in Section \ref{sec:termination} we terminate the simulation once the accretion timescale approaches the stellar dynamical timescale.

Fig.~\ref{fig:demo_radius_and_mass} also demonstrates that the dissipative forces lead to a shrinking of the orbit, so that the PBH sinks toward the center of the host star as expected.  The figure further shows that the ratio between the pericenter and apcenter distances hardly changes during this process, so that the orbit's eccentricity does not change significantly - which we will explore in greater detail in Section \ref{sec:results:eccentricity} below.  Finally, we note that the PBH's orbit precesses, which is a consequence of both inhomogeneity and relativistic effects (see \cite{BauS24b}).  This precession results in beat-like modulations of the gravitational-wave signal, as we discuss in more detail in Section \ref{sec:results:GW_wave} below.

%
\subsection{Capture and confinement}
\label{sec:results:capture}
%

\begin{figure}
    \includegraphics[width = .45 \textwidth]{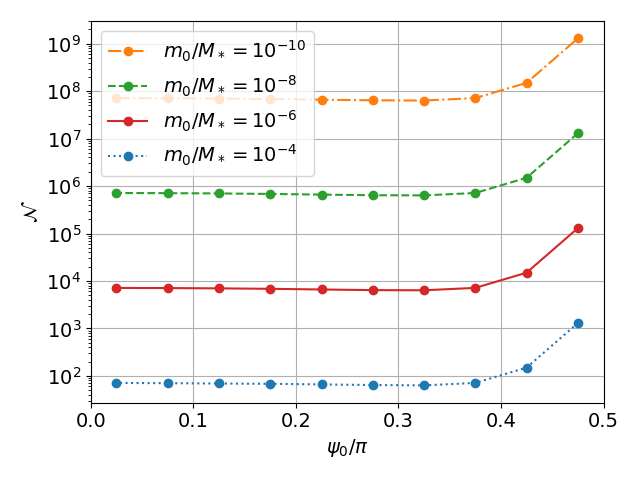}  
    \caption{Number of orbits ${\mathcal N}$ for different initial black-hole masses $m(0)$ and initial impact angles $\psi_0$, where $\psi_0 = 0$ corresponds to a head-on collision and $\psi_0 = \pi/2$ to a grazing collision.  As expected from the estimate (\ref{N_estimate}) we observe ${\mathcal N} \propto M_*/m(0)$.}
    \label{fig:captureN}
\end{figure}

\begin{figure}
    \includegraphics[width = .45 \textwidth]{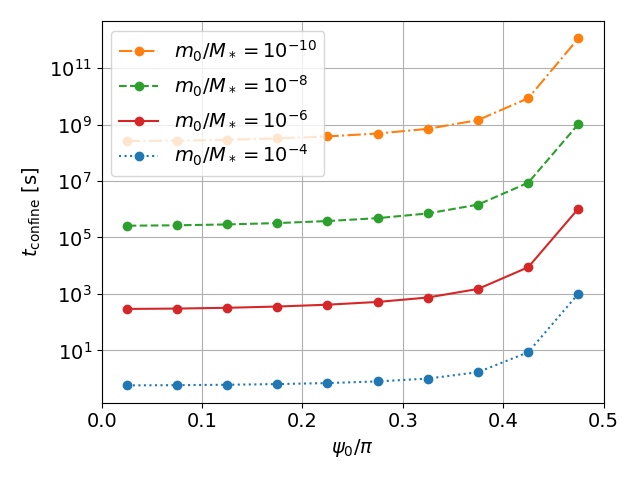}
    \caption{Same as Fig.~\ref{fig:captureN}, except that we plot the times $t_{\rm confine}$ from the first passage until the PBH no longer emerges from the neutron star.  We caution that these times are dominated by the first few orbits and depend sensitively on the speed $v_\infty$ (chosen to be 1 km/s here); the numbers shown therefore serve as examples only.}
    \label{fig:capturetime}
\end{figure}

We next turn to smaller PBH masses in the range $10^{-10} \leq m(0) / M_* \leq 10^{-4}$.  In Fig.~\ref{fig:captureN} we show the number of orbits ${\mathcal N}$ that are required until the PBH is completely absorbed by the host star, and in Fig.~\ref{fig:capturetime} the corresponding confinement times $t_{\rm confine}$.  We observe that ${\mathcal N}$ is proportional to the mass ratio $M_*/m(0)$, as expected from the estimate (\ref{N_estimate}).  The times $t_{\rm confine}$ are dominated by the first few orbits, when the orbital energy is close to zero and the semi-major axis very large.  The initial energy, and hence the orbital period, depends very sensitively on the speed at infinite separation $v_\infty$, so that the times shown in Fig.~\ref{fig:capturetime} should be taken as illustrative examples only.

\begin{figure}
    \includegraphics[width = .45 \textwidth]{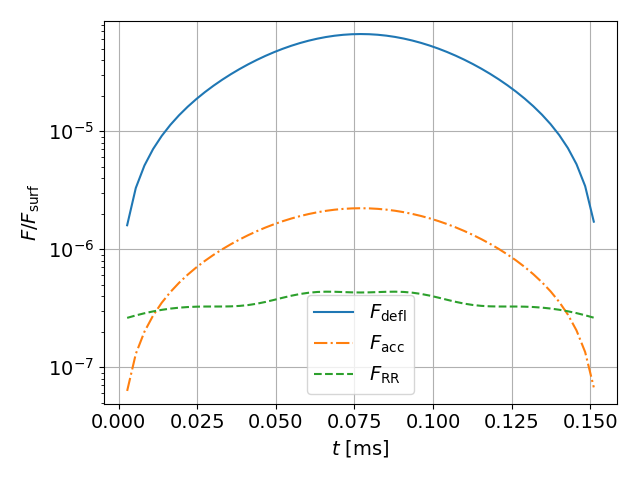}
    \caption{The magnitudes of the deflection force $F_{\rm defl}$, the accretion force $F_{\rm acc}$, and the radiation-reaction force $F_{\rm RR}$ during the initial passage through the star, for $m(0) = 10^{-6} M_*$ and $\psi_0 = 0.707$.  We show the fractions between the above forces and the Newtonian gravitational force $F_{\rm surf} = G M_* m(0)/r_*^2$ that the PBH would experience at the stellar surface.}
    \label{fig:initialforces}
\end{figure}

\begin{figure}
    \includegraphics[width = .45 \textwidth]{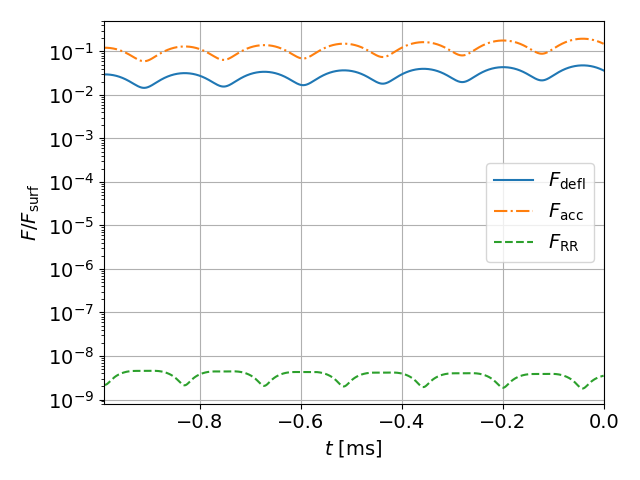}
    \caption{Same as Fig.~\ref{fig:initialforces}, except that we show the forces during the last few orbits, just before we terminate the calculation (taken to be at time $t = 0$ here).  Unlike at early times, when the dissipative forces are dominated by the deflection force, they are dominated by the accretion force at late times.}
    \label{fig:finalforces}
\end{figure}

In Fig.~\ref{fig:initialforces} we show the three different dissipative forces in (\ref{F_diss}), i.e.~the deflection force $F_{\rm defl}$, the accretion force $F_{\rm acc}$, and the radiation-reaction force $F_{\rm RR}$, during the first transit through the host star for a PBH with $m(0) = 10^{-6} M_*$ and an initial penetration angle of $\psi_0 = 0.707$.  We show each force in units of $F_{\rm surf} = G M_* m(0)/r_*^2$, i.e.~the Newtonian gravitational force that the PBH would experience at the surface of the neutron star.  Evidently, the deflection force dominates during this first transit, justifying our assumptions in Section \ref{sec:timescales}.  Note also that the initial transit takes a time of about 0.15~ms, and hence similar to the dynamical timescale $\tau_{\rm dyn} \simeq 0.1$~ms (see \ref{tau_dyn}), as expected from the estimate (\ref{transit_time}).  

In Fig.~\ref{fig:finalforces} we show the same forces as in Fig.~\ref{fig:initialforces} for the last few orbits, demonstrating that, at late times, the accretion force dominates over dynamical friction.

%
\subsection{Accretion and collapse}
\label{sec:results:accretion}
%

\begin{figure}
    \includegraphics[width = .45 \textwidth]{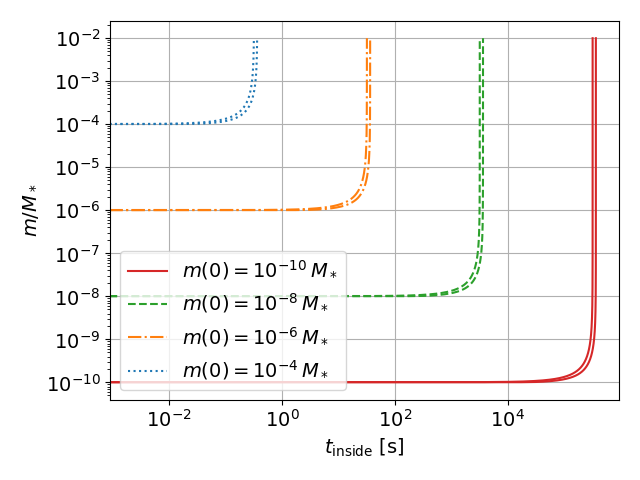}
    \caption{The PBH mas $m$ as a function of time $t_{\rm inside}$, i.e.~the time after which the PBH no longer emerges from the neutron star, for different initial PBH masses $m(0)$.  For each initial mass we include results for two different initial impact angles $\psi_0$, namely for a nearly head-on collision with $\psi_0 = \pi / 40$, which leads to slightly shorter collapse times, and for a nearly grazing capture with $\psi_0 = 19 \pi/40$, which leads to slightly longer collapse times. }
    \label{fig:accretion}
\end{figure}

We next discuss the accretion process once the PBH is completely contained within the neutron star.  In Fig.~\ref{fig:accretion} we show the PBH mass $m$ as a function of time after it no longer emerges from the star, which we refer to as $t_{\rm inside}$.  As a result of the accretion rate $\dot m$ being proportional to $m^2$ (see Eq.~\ref{bondi}), it increases very slowly initially, when $m$ is small, and very rapidly late, when $m$ is large.  This faster-than-exponential growth is clearly visible in the log-log plot in Fig.~\ref{fig:accretion}.  We terminate our simulations when $m = 0.01 M_*$ (see Section \ref{sec:termination}), but given the rapid growth of $m$ at these late times the precise value of this collapse criterion has only an extremely small effect on the resulting collapse time.  We note that the collapse time $t_{\rm coll}$, i.e.~the times spent inside the star until the accretion process triggers collapse, are proportional to $m(0)^{-1}$.  Further using the dynamical timescale (\ref{tau_dyn}) we have
\begin{equation}
t_{\rm coll} \simeq \frac{M_*}{m(0)} \, \tau_{\rm dyn},
\end{equation}
in accordance with our estimate (\ref{tau_accretion}) for the accretion timescale $\tau_{\rm acc}$.

Based on the minimum accretion rate discussed in Section \ref{sec:accretion} we previously estimated a {\em maximum survival} time for neutron stars hosting an endoparasitic black hole,
\begin{equation}
t_{\rm coll}^{\rm max} \simeq 10^{6}\,\mbox{s}\, \left(\frac{10^{-10} M_\odot}{m(0)} \right)
\end{equation}
(see Eq.~17 in \cite{BauS21}).  Using $M_* = 1.4 M_\odot$ for our choices here we then obtain
\begin{equation} \label{t_coll_max}
t_{\rm coll}^{\rm max} \simeq 7 \times 10^{5}\,\mbox{s}\, \left(\frac{10^{-10} M_*}{m(0)} \right).
\end{equation}
All the collapse times found in Fig.~\ref{fig:accretion} are similar but smaller than the maximum survival times (\ref{t_coll_max}), as expected, even though our estimates in \cite{BauS21} assumed the PBH to reside at the stellar center and did not take into account any motion of the PBH through the star.

Finally we note that the confinement times $t_{\rm confine}$, shown in Fig.~\ref{fig:capturetime}, are significantly longer than the collapse times $t_{\rm coll} \simeq \tau_{\rm acc}$, at least for small initial PBH masses, confirming our estimate (\ref{tau_hierarchy}).

%
\subsection{Gravitational-wave signal}
\label{sec:results:GW_wave}
%

\begin{figure}
    \includegraphics[width = .5 \textwidth]{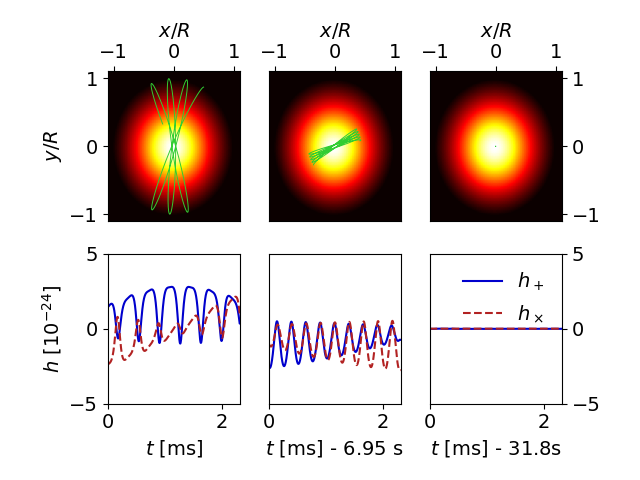}
    \caption{Snapshots of the orbit and emitted gravitational-wave signal for the capture of a PBH with $m(0) = 10^{-6} M_*$ and for a nearly head-on collision with $\psi_0 = \pi/40 = 0.07854$.  The left column shows the first few orbits after the PBH no longer emerges from the neutron star, the middle column at an intermediate time, and the right column shortly before the star becomes unstable and we terminate the simulation.   The orbit of the PBH, shown as the green lines in the top panels, precesses in the counter-clockwise direction; the color coding in these panels is the same as in Fig.~\ref{fig:demo_orbit}.  The two gravitational-wave polarizations $h_+$ and $h_\times$, as emitted in the direction of the $z$-axis, are scaled assuming a distance of 10 kpc to the neutron star.}
    \label{fig:composite_headon}
\end{figure}

\begin{figure}
    \includegraphics[width = .5 \textwidth]{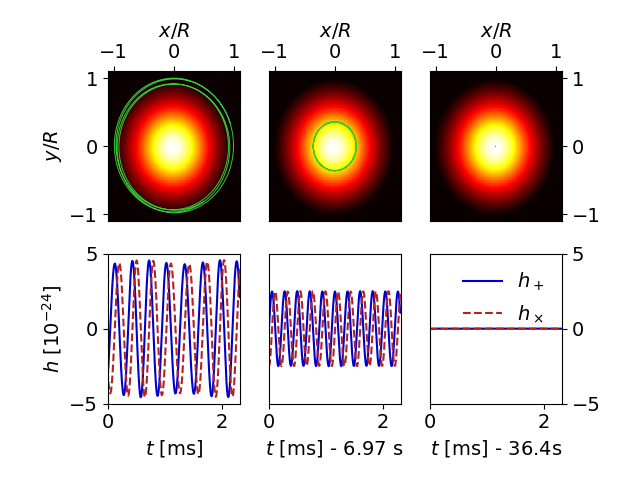}
    \caption{Same as Fig.~\ref{fig:composite_headon}, except for a nearly grazing collision with $\psi_0 = 1.492$, which results in a nearly circular orbit.}
    \label{fig:composite_grazing}
\end{figure}

The gravitational wave signal emitted by PBHs confined inside neutrons, and the prospect of detecting such signals with either current or next-generation gravitational-wave detectors, has previously been discussed by a number of authors, including \cite{HorR19,GenST20,ZouH22,GaoDGZZZ23,BauS24b}.  We note that \cite{ZouH22,GaoDGZZZ23} considered spherical orbits only, which appears unrealistic in light of our findings of Section \ref{sec:results:eccentricity} below.  As before we consider here orbits with different eccentricities, resulting from different initial impact angles $\psi_0$. 

In Figs.~\ref{fig:composite_headon} and \ref{fig:composite_grazing} we show snapshots of orbits and gravitational-wave signals for the capture of a PBH with $m(0) = 10^{-6} M_*$, both for a nearly head-on collision (Fig.~\ref{fig:composite_headon}) and a nearly grazing collision (Fig.~\ref{fig:composite_grazing}).  

Focusing on the orbits first, shown in the top rows of the two figures, we see again that the PBH slowly drifts toward the stellar center.  As anticipated from Fig.~\ref{fig:demo_radius_and_mass}, and as discussed in more detail in Section \ref{sec:results:eccentricity} below, the eccentricity does not decrease significantly during this process.  We further observe a pericenter advance for noncircular orbits.  As we discussed in \cite{BauS24b}, this pericenter advance results from inhomogeneity even in Newtonian gravity, but is dominated by relativistic effects.  Since the gravitational-wave signal is dominated by the $h_+$ polarization when the orbit aligns with either the $x$ or the $y$-axis, but by the $h_\times$ polarization when it is aligned with the diagonals, this pericenter advance results in beats in the each one of the gravitational-wave polarizations.  The pericenter advance and hence the beat frequency depend sensitively on the stellar structure, so that a future detection of these signals by next-generation gravitational-wave detectors would provide constraints on the neutron-star equation of state (see \cite{BauS24b} for details). 

The characteristics of the gravitational-wave signal emitted by the capture orbit of a PBH inside a neutron star are quite different from those emitted by other common or promising sources.  In a typical {\em chirp signal}, for example, emitted by a binary inspiral, both the gravitational-wave amplitude and frequency increase with time.  For stellar-mass binaries, these sources also spend only a short time in the frequency range in which ground-based gravitational-wave detectors are most sensitive.   A promising source of {\em continuous} gravitational wave signals are spinning neutron stars, for which both the amplitude and frequency decrease slowly (see, e.g., \cite{Ril23} for a recent review).  The capture orbits considered here also represent a source of continuous gravitational waves, but its features are qualitatively different from those emitted by a spinning neutron star.   First, we observe that the gravitational-wave frequency increases slightly early on, and then remains approximately constant.  This is to be expected from a simple Newtonian argument: Approximating the stellar density to be constant (which becomes increasingly accurate as the PBH sinks towards the stellar center), the orbital frequency is given by
\begin{equation}
\Omega^2 \simeq \frac{M(r)}{r^3} \simeq \rho
\end{equation}
and hence approaches a constant value (related to the dynamical time-scale $\tau_{\rm dyn}$) in the neutron star core.  Similarly, the gravitational-wave amplitude $h$ decreases during the capture process, which we can anticipate from the crude dimensional estimate
\begin{equation}
h \simeq \ddot {\mathcal I} \simeq
\frac{m r^2}{\tau_{\rm dyn}^2} \propto m r^2
\end{equation}
where ${\mathcal I}$ represents the reduced quadrupole moment (see Eq.~\ref{quadrupole} in Appendix \ref{sec:radreac:general}), and where we have used the previous result that $\tau_{\rm dyn} \simeq \Omega^{-1}$ remains approximately constant.  While $m$ increases during evolution, it increases more slowly than $r^2$ decreases, so that the product $m r^2$ approaches zero at late times.   Superimposed on this signal is an amplitude modulation caused by the precession of the orbit as we discussed above; the frequency of the resulting gravitational-wave beats decreases as the orbit shrinks.

%
\subsection{Eccentricity and semi-major axis}
\label{sec:results:eccentricity}
%

\begin{figure}
    \includegraphics[width = .45 \textwidth]{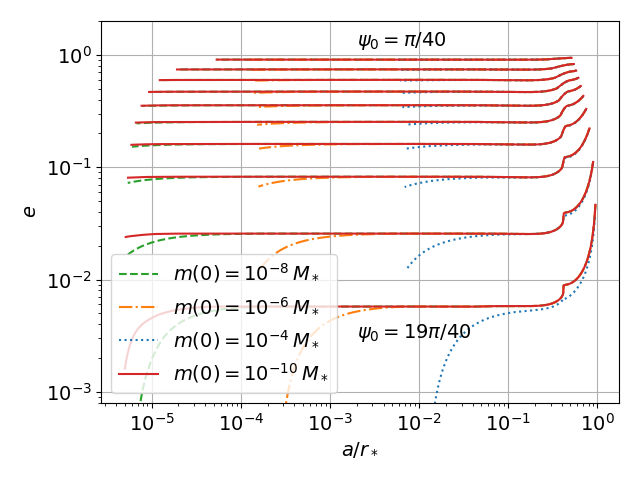}
    \caption{Evolutionary trajectories in an eccentricity $e$ versus semi-major axis $a$ diagram after the black hole is completely contained in the host star.  Each group of lines corresponds to a specific initial impact angle $\psi_0$, increased by increments of $\pi / 20$; the top lines with the highest eccentricity correspond to our smallest values of $\psi_0 = \pi / 40$, close to a head-on collision, while the bottom lines correspond to our shallowest initial passages with $\psi_0 = 19 \pi/40$, close to a grazing collision.   All trajectories start at the right, with large values of the semi-major axis right after the black hole is completely swallowed by the host star, and proceed towards the left, where they end when the black-hole mass exceeds $\m = 10^{-2}$, shortly before the host star would collapse dynamically. }
    \label{fig:eccentricity}
\end{figure}

\begin{figure}
    \includegraphics[width = .45 \textwidth]{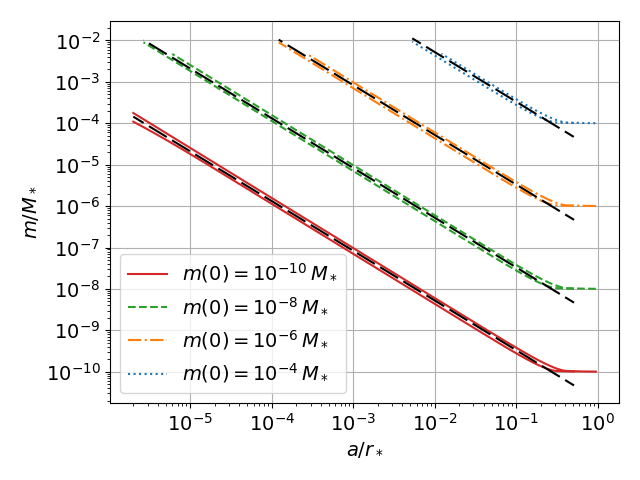}
    \caption{Similar to Fig.~\ref{fig:accretion}, except that we show the PBH mass as a function of semi-major axis $a$.  For each initial PBH mass $m(0)$ the lower line corresponds to $\psi_0 = \pi / 40$, and the upper line to $\psi_0 = 19 \pi/40$.  The black long-dashed lines are the fits (\ref{m_versus_a_fit}).}
    \label{fig:m_versus_a}
\end{figure}

We have previously observed that the eccentricity of the PBH's orbit inside the neutron star does not appear to decrease substantially, and we now examine this behavior more systematically.  Specifically, we track the pericenter and apcenter distances $r_{\rm p}$ and $r_{\rm ap}$ once the PBH has been completely absorbed by the neutron star, and, from these, adopt Newtonian expressions\footnote{We note that the PBH's orbit inside the star is {\em not} a Keplerian orbit.  In fact, in Newtonian gravity and in the limit of constant density, e.g.~in the vicinity of the center, the orbit approaches a nonprecessing closed orbit with two pericenters per orbit; see \cite{BauS24b} for a discussion.} to compute the semi-major axis $a$
\begin{equation} \label{semimajor}
a \equiv\frac{r_{\rm ap} + r_{\rm p}}{2}
\end{equation}
and the eccentricity $e$
\begin{equation}
    e \equiv \frac{r_{\rm ap} - r_{\rm p}}{r_{\rm ap} + r_{\rm p}}
\end{equation}
of orbits in the stellar interior.\footnote{Recall that the expressions (\ref{ae_Newton}) apply in the stellar exterior only.}  In Fig.~\ref{fig:eccentricity} we graph $e$ versus $a$ for the same initial data as those in Figs.~\ref{fig:captureN} and \ref{fig:capturetime}, i.e.~for different values of the initial PBH mass $m(0)$ and the initial impact angle $\psi_0$.  All evolutionary tracks in Fig.~\ref{fig:eccentricity} start at the right, for large semi-major axis $a$, and terminate further left at a smaller value of $a$.

Consistent with our earlier observations Fig.~\ref{fig:eccentricity} demonstrates that the eccentricity $e$ decreases only by small amounts while the PBH sinks to the stellar center, especially for orbits that start out with a large eccentricity.  There is a small decrease in $e$ early on that, for small initial $e$, appears rather abrupt in Fig.~\ref{fig:eccentricity}.  This sharp transition is related to the transition from supersonic to subsonic flow.  The eccentricity $e$ decreases again late in the inspiral once the mass $m$ and the accretion rate $\dot m$ become large.  For most of the evolution, however, the eccentricity changes very little in our simulations.  We therefore conclude that the dissipative forces are not effective in circularizing the PBH orbit, at least for the models considered here (see also the discussion in \cite{Hof85,SzoML22}).

In Fig.~\ref{fig:m_versus_a} we show the 
growing PBH masses $m$ as a function of the orbit's shrinking semimajor axis $a$, as defined in Eq.~(\ref{semimajor}).  From the plot it is evident that, after a brief initial transition, the relation between $a$ and $m$ is well described by a power law and depends on the initial impact angle $\psi_0$ only weakly.  We found that all lines are well approximated by the fit
\begin{equation} \label{m_versus_a_fit}
m = 0.045 \left( \frac{m(0)}{M_*} \right) \left(\frac{a}{r_*} \right)^{-1.2},
\end{equation}
which we included as the black long-dashed lines in Fig.~\ref{fig:m_versus_a}.  While the power-law behavior can be motivated from a simple Newtonian argument (see Appendix \ref{sec:m_versus_a}), the specific value of the exponent (here close to -1.2) depends on the structure and compaction of the neutron star, as well as effects of general relativity.

Some previous authors (e.g.~\cite{ZouH22,GaoDGZZZ23}) have considered gravitational-wave signals from PBH with initial masses $m(0) \geq 10^{-6} M_\odot$ in circular orbits inside neutrons stars, starting either at or close to the stellar radius.  Our observations above suggest that such orbits could result only from a near-grazing capture of a PBH, which starts out with near-zero eccentricity (see Fig.~\ref{fig:eccentricity}).  By contrast, it appears unlikely that such an orbit could result from the circularization of a more eccentric initial orbit.   We also note that a captured PBH that starts out with a smaller mass $m(0)$ will ultimately grow to reach $m \geq 10^{-6} M_\odot$, but will, in the meantime, sink to smaller radii inside the neutron star (see Fig.~\ref{fig:m_versus_a}).

%
\subsection{Effects of dissipation}
\label{sec:results:dissipation}
%

\begin{figure}
    \includegraphics[width = .45 \textwidth]{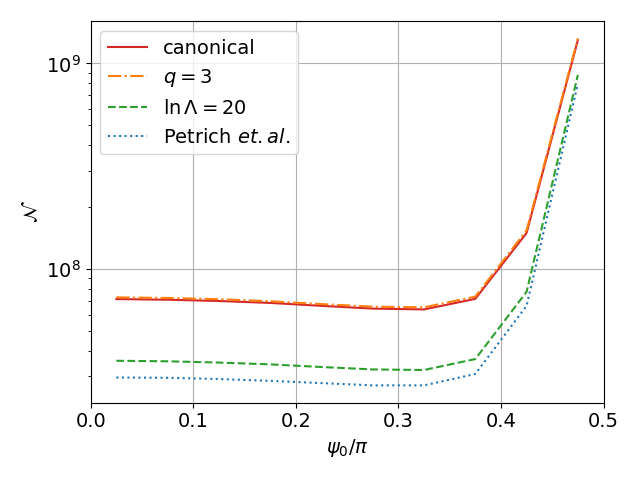}  
    \caption{Similar to Fig.~\ref{fig:captureN}, except that we focus on $m(0) = 10^{-10} M_*$ here and compare our ``canonical" choices for the dissipative forces with other choices (see text for details). }
    \label{fig:compareN}
\end{figure}

\begin{table}[]
    \centering
    \begin{tabular}{c|c|c|c}
         & ${\mathcal N}~[10^7]$ & $t_{\rm confine}~[10^8$ s] &  $t_{\rm coll}~[10^5$ s]  \\
         \hline 
         Canonical & 7.13 & 2.61 & 3.18 \\
         $q = 3$ & 7.27 & 2.68 & 3.49 \\
         $\ln \Lambda = 20$ & 3.57 & 0.92 & 3.17 \\ 
        Petrich {\it et.al.} & 2.96 & 0.69 & 3.47
    \end{tabular}
    \caption{Comparison of capture data for different treatments of the dissipative forces.  We consider a PBH with initial mass $m(0) = 10^{-10} M_*$, initially striking the host star in an almost head-on collision with an impact angle of $\psi_0 = \pi / 40$, and compare the number ${\mathcal N}$ of orbits completed before PBH is completely absorbed by the neutron star and no longer re-emerges, the capture time $t_{\rm capt}$ required for this process, and the additional collapse time $t_{\rm coll}$ that passes before the PBH triggers the gravitational collapse of the neutron star. }
    \label{tab:capture}
\end{table}

For all previous results we adopted the ``canonical" choices for the dissipative forces discussed in Section \ref{sec:dissipation}, namely the expressions (\ref{dyn_friction_ostriker}) of Ostriker \cite{Ost99} for the dynamical friction with $\ln \Lambda = 10$ and relativistic accretion with $q = 2$ in (\ref{f_accretion_final}).   In this section we now examine effects of these choices by comparing with some alternatives.  Specifically, we consider choosing an exponent $q = 3$ rather than $q =2$ in the accretion drag (\ref{f_accretion_final}), choosing $\ln \Lambda = 20$ rather than $\ln \Lambda = 10$ in Ostriker's dynamical friction (\ref{dyn_friction_ostriker}), and, finally, adopting the prescription (\ref{dyn_friction_petrich}) of Petrich {\em et.al.} \cite{PetSST89} for the dynamical friction.   

In Fig.~\ref{fig:compareN}, which is similar to Fig.~\ref{fig:captureN}, we show the number of orbits ${\mathcal N}$ completed until the PBH no longer emerges from the neutron star.  Unlike in Fig.~\ref{fig:captureN} we focus on $m(0) = 10^{-10} M_*$ here, but show results for both the canonical and alternative choices.   We observe that the choice of $q$ 
in the accretion rate hardly affects ${\mathcal N}$ at all, which is consistent with our previous finding that, at early times, the dissipative forces are dominated by dynamical friction.  Accordingly, a larger value of $\ln \Lambda$ should lead to smaller values of ${\mathcal N}$, which is what we observe.  Using (\ref{Petrich_Lambda}) to estimate $\ln \Lambda$ in the prescription of Petrich {\em et.al} typically results in values that are closer to 20 rather than 10 for supersonic flow, so that ${\mathcal N}$ is again smaller than the canonical values using this treatment. 

\begin{figure}
    \includegraphics[width = .45 \textwidth]{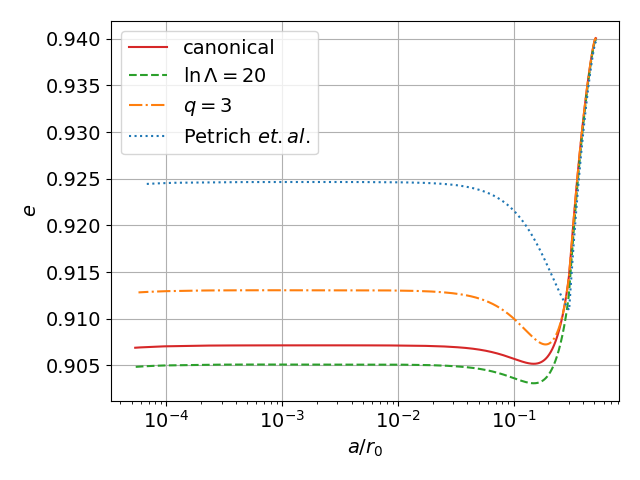}
    \caption{Similar to Fig.~\ref{fig:eccentricity}, except that we focus on $m(0) = 10^{-10} M_*$ and a highly eccentric orbit with $\psi_0 = \pi/40$ here and compare the effects of the choices for the dissipative forces on the evolution of the eccentricity.}
    \label{fig:comparee}
\end{figure}

In Table \ref{tab:capture} we provide numerical values for ${\mathcal N}$, the capture times $t_{\rm capt}$, and the collapse times $t_{\rm coll}$ for $m(0) = 10^{-10} M_*$ and a near head-on capture with $\psi_0 = \pi / 40$.  Both ${\mathcal N}$ and $t_{\rm capt}$ behave according to the expections that we discussed above.  The collapse time $t_{\rm coll}$, which measures the time starting when the PBH no longer emerges from the neutron star and ending when it induces the dynamical gravitational collapse of the host star, depends only weakly on the choices of the dissipative forces.  For a larger value of $q$ in (\ref{f_accretion_final}) the accretion rate is suppressed more, meaning that the PBH grows more slowly,  resulting in a longer collapse time.  However, once the PBH slows down and its speed becomes well subsonic, the effect of this suppression term is small.  In the Petrich {\it et.al.}~prescription the dynamical friction for subsonic speed vanishes (see \ref{dyn_friction_petrich}), which means that the PBH will maintain a higher speed for longer, meaning that the accretion rate is suppressed for longer, which also results in a slightly longer collapse time than in the canonical case.  We also note that all the collapse times listed in Table \ref{tab:capture} are again somewhat smaller than the maximum survival time (\ref{t_coll_max}).  The key result is that the values for these timescales yield about the same magnitude within a factor of order unity for the different choices adopted for the dissipative forces

Finally we examine effects of the prescriptions for the dissipative forces on the evolution of the eccentricity $e$.  Focusing on highly eccentric orbits with $m(0) = 10^{-10} M_*$, we show $e$ versus $a$ both for our canonical choice and alternatives in Fig.~\ref{fig:comparee} (compare Fig.~\ref{fig:eccentricity}).  While different choices lead to some small quantitative differences, our qualitative finding that the eccentricity does not decrease significantly appears to be independent of the detailed treatment of the dissipative forces.

%
\section{Summary and Discussion}
\label{sec:discussion}
%

We consider the gravitational capture and confinement of a PBH by a neutron star, leading to the formation of a TZlO.  We follow the co-evolution of the PBH and the neutron star, as the PBH orbits inside the host toward its center and grows in mass as it accretes stellar material.  In this paper we adopt a relativistic point-mass treatment that is suitable for sufficiently small PBH masses so that the effect of the PBH on the neutron star can be ignored. 

We consider collisions between PBHs and neutron stars for different PBH masses and initial impact angles and find good agreement with the estimates presented in Section \ref{sec:timescales}.  We examine the gravitational-wave signal emitted while the PBH orbits inside the host star, and 
contrast its characteristics with those from other common or promising sources.  For a typical chirp signal emitted by a stellar-mass binary inspiral both the frequency and amplitude of the gravitational-wave signal increase rapidly in the frequency range of ground-based detectors.  Promising sources of continuous gravitational waves are spinning neutron stars, for which both amplitude and frequency slowly decrease.  For the capture orbits considered here, the frequency is expected to increase slightly early on before settling down to a nearly constant value related to the host star's dynamical timescale, while the amplitude decreases.  Moreover, the amplitude is modulated by a precession of the orbit, which leads to quasiperiodic beats whose frequency depends sensitively on the neutron star structure and hence the stiffness of its equation of state.  A future observation of these beats may therefore provide strong constraints on the nuclear equation of state (see \cite{BauS24b}).   

We find that the dissipative forces acting on the PBH inside the neutron star are quite ineffective in circularizing its orbit, so that the orbit's eccentricity does not decrease significantly.  This result is important in light of the beat frequencies discussed above, because it means that the size of the beats, i.e.~the degree of the gravitational-wave amplitude modulation, will persist throughout the capture orbit.

Finally, we explore the effects of several choices in the prescription of the dissipative forces acting on the PBH.  As expected, these choices lead to some small quantitative differences, but do not affect our findings qualitatively. 

As we stated above, our point-mass treatment adopted here assumes that the effects of the PBH on the neutron star can be neglected.  This assumption holds while the PBH mass is small compared to that of the neutron star, but breaks down either late in the evolution, after the PBH has accreted a significant fraction of the neutron star mass, or for PBHs with larger initial masses than those considered here. In  \cite{BauS24c} we therefore relax this assumption and perform fully self-consistent dynamical simulations in order to ascertain when a neutron star can survive the passage of a PBH through its center, and to study the effects of larger PBH masses on the neutron star structure and stability.


\acknowledgments

We would like to thank Keith Riles for helpful conversations. T.W.B.~gratefully acknowledges hospitality at the University of Illinois at Urbana-Champaign's Center for Advanced Studies of the Universe.  This work was supported in part by National Science Foundation (NSF) grants PHY-2010394 and PHY-2341984 to Bowdoin College, as well as NSF grants PHY-2006066 and PHY-2308242 to the University of Illinois at Urbana-Champaign.

\begin{appendix}

%
\section{Exterior Motion}
\label{sec:exterior}
%

As discussed in Section \ref{sec:numerics}, we treat the PBH's motion in the stellar exterior in a Newtonian framework, ignoring all effects of relativistic gravity and radiation reaction, so that its trajectory is given by a segment of a Keplerian orbit.

Specifically, we identify Newtonian values of the energy $E_{\rm Newt}$ and angular momentum $L_{\rm Newt}$ per unit mass, see (\ref{EL_Newton}), and, from these, compute the orbit's semi-major axis $a$ and eccentricity $e$, see (\ref{ae_Newton}).  We next equate $E_{\rm Newt}$ with the Newtonian expression for the total energy
\begin{equation}
E_{\rm Newt} = \frac{1}{2} (v^r)^2 + \frac{L_{\rm Newt}^2}{2 r^2} - \frac{M}{r}
\end{equation}
and solve for $v^r$ to find 
\begin{align} \label{vr}
v^r & = \left( \frac{2 E_{\rm Newt}}{r^2} \left( r^2 + \frac{M r}{E} - \frac{L_{\rm Newt}^2}{2 E} \right) \right)^{1/2}  \nonumber \\
& = \left( \frac{2 |E_{\rm Newt}|}{r^2} (r_{\rm ap} - r) (r - r_{\rm p}) \right)^{1/2},
\end{align}
where
\begin{equation}
r_{\rm ap} = a (1 + e),~~~~~
r_{\rm p} = a (1 - e)
\end{equation}
are the apcenter and pericenter distances, respectively.

We can now find the time $\Delta t$ between emergence from the host star and re-entry from
\begin{equation} \label{deltat1}
\Delta t = 2 \int_{r_*}^{r_{\rm ap}} \frac{d r}{v^r}.
\end{equation}
Using (\ref{vr}) we can rewrite this integral as
\begin{equation} \label{deltat2}
\Delta t = \left( \frac{2}{|E_{\rm Newt}|} \right)^{1/2} \int_{r_*}^{r_{\rm ap}} \frac{r dr}{\left( (r_{\rm ap} - r) (r - r_{\rm p}) \right)^{1/2}},
\end{equation}
which can be integrated in closed form using the trig substitution
\begin{equation}
r = \frac{r_{\rm ap} + r_{\rm p}}{2} + \frac{r_{\rm ap} - r_{\rm p}}{2} \sin \theta = a + e a \sin \theta.
\end{equation}
The result of the integration is
\begin{align} \label{deltat3}
\Delta t = \left( \frac{2}{|E_{\rm Newt}|} \right)^{1/2} \Big[ & a \arcsin \left( \frac{r- a}{a e}\right) \nonumber \\ 
& - \left( (r_{\rm ap} - r)(r- r_{\rm p}) \right)^{1/2} \Big]^{r_{\rm ap}}_{r_*}.
\end{align}
Inserting the limits we obtain\footnote{In some cases the identification of the Newtonian energy and angular momentum from their relativistic counterparts leads to an estimate for $r_{\rm p}$ that is slightly larger than $r_*$.  In that case we replace $r_*$ with $r_{\rm p}$ in Eq.~(\ref{deltat3}) and (\ref{deltat4}), which amounts to computing the period of the entire orbit rather than ignoring the fraction of the orbital time that would otherwise be spent inside the star.  Since the latter is small, the effect on our overall estimates is also small.} 
\begin{align} \label{deltat4}
\Delta t = \left( \frac{2}{|E_{\rm Newt}|} \right)^{1/2} \Big\{ &
(2a r_* - r_*^2 - r_{\rm ap} r_{\rm p} )^{1/2} \nonumber \\
& + a \frac{\pi}{2}
- a \arcsin \left( \frac{r_* - a}{a e} \right) \Big\}.
\end{align}

%
\section{Effects of gravitational radiation}
\label{sec:radreac}
%

In this Appendix we develop all expressions needed to compute the leading-order effects of gravitational-wave (GW) radiation reaction
on the orbital motion, including the associated losses of energy and angular momentum, and the lowest-order GW strains.  In Section \ref{sec:radreac:general} we derive the radiation-reaction force acting on the black hole, both in the interior and exterior of the star, in Section \ref{sec:radreac:strain} we compute the gravitational-wave strains, and in Section \ref{sec:radreac:exterior} we integrate the radiation-reaction forces in the stellar exterior to obtain analytical expressions for the energy and angular momentum lost in the orbital segments outside the star (in the stellar interior, we evaluate these losses numerically).  Our approach applies to slow-motion, weak-field systems and thereby adopts the quadrupole formalism adopted by numerous previous authors (see, e.g., \cite{MisTW73} for a textbook treatment, as well as \cite{Bur71} for references to early work on this topic), and we only take into account the lowest order Newtonian terms in the PBH's equations of motion.  

%
\subsection{Radiation-reaction force}
\label{sec:radreac:general}
%

To leading order, the radiation-reaction force acting on a point-mass $m$ can be expressed as 
\begin{equation} \label{AFRR0}
    F_i^{\rm RR} = - m \nabla_i \Phi^{\rm RR}.
\end{equation}
Here the radiation-reaction potential
\begin{equation} \label{APhiRR}
\Phi^{\rm RR} = \frac{G}{5c^5} \I_{jk}^{(5)} x^j x^k
\end{equation}
is given in terms of the fifth time derivative of the reduced quadrupole moment 
\begin{equation} \label{quadrupole}
\I_{jk} = m \left( x_j x_k - \frac{1}{3} \delta_{jk} r^2 \right).
\end{equation}
We include the gravitational constant $G$ and the speed of light $c$ in most equations of this appendix for clarity, we adopt the Einstein summation convention in (\ref{APhiRR}) and in the following, by which repeated indices are summed over, and a supercript $(n)$ denotes the $n$-th derivative with respect to time $t$.  Using (\ref{APhiRR}) we may also write the radiation-reaction force (\ref{AFRR0}) as
\begin{align} \label{AFRR0a}
    F_i^{\rm RR} = - \frac{2Gm}{5c^5} \,\I_{ij}^{(5)} x^j. 
\end{align}

We adopt Cartesian coordinates (for which we do not need to distinguish between upper and lower indices) and assume that the point mass orbits in the $z=0$ plane, in which case the only non-vanishing components of $\I_{jk}$ are
\begin{subequations} \label{AI}
\begin{align}
\I_{xx} & = \frac{m}{3} \left( 2 x^2 -  y^2 \right) \\
\I_{xy} & = m  x y \\
\I_{yy} & = \frac{m}{3} \left( 2 y^2 -  x^2 \right) \\
\I_{zz} & = - \frac{m}{3} \left( x^2 + y^2 \right) 
\end{align}
\end{subequations}
We next compute the first five time derivatives of (\ref{AI}), replacing second time derivatives of the coordinates $x$ and $y$ with the leading-order terms in the equations of motion
\begin{subequations} \label{AEOM}
\begin{align}
\ddot x = - \frac{G M(r)}{r^3} x \\
\ddot y = - \frac{G M(r)}{r^3} y
\end{align}
\end{subequations}
(with $r = (x^2 + y^2)^{1/2}$) whenever they occur.  We note that we have assumed the host star to be fixed and centered on the origin in (\ref{AEOM}), which is consistent with our approximation $m \ll M$ that we will adopt throughout.  Inserting the moments into (\ref{AFRR0a}) yields the Cartesian components of the radiation-reaction force, which we then transform into polar coordinates to obtain
\begin{subequations} \label{AFRR1}
\begin{align}
& F^r_{\rm RR} =  \frac{8 G^2 m^2}{15 c^5 r^4} \dot r \Big\{ 8 G M(r)^2 \\
& ~~~~~~ + r M(r) \Big( 6 \dot r^2 + 36 r^2 \dot \varphi^2 +
G M'(r) - 3 r G M''(r) \Big) \nonumber \\
& ~~~~~~ - r^2 \Big(G M'(r)^2 + 3 r^2 \dot \varphi^2(6 M'(r) - r M''(r)) \nonumber \\ 
& 
~~~~~~ ~~~~~~~~ +\dot r^2 \big( 2 M'(r) + r M''(r) - r^2 M'''(r) \big) \Big) \Big\} \nonumber 
\end{align}
and
\begin{align} \label{AFRRphi}
F^\varphi_{\rm RR} = ~ & \frac{8 G^2 m^2}{5 c^5 r^4} \dot \varphi \Big\{2 G M(r)^2 + 2 r^4 \dot \varphi^2 M'(r) \\
& + r M(r) \Big(9 \dot r^2 
- 6 r^2 \dot \varphi^2 - 2 G M'(r) \Big) \nonumber \\
& - r^2 \dot r^2 \big( 7 M'(r) - 2 r M''(r) \big) \Big\}. \nonumber
\end{align}
\end{subequations}
Here a prime on the enclosed mass $M(r)$ denotes a derivative with respect to radius $r$.  


We can also compute the change in the point-mass' energy from
\begin{align} \label{AdEdt2a}
\frac{dE}{dt} & = v_i F^i_{\rm RR}
= - \frac{2 G m}{5 c^5} v^i x^j \I^{(5)}_{ij} = - \frac{G}{5 c^5} \frac{d}{dt}(m x^i x^j) \I^{(5)}_{ij} \nonumber \\
& = - \frac{G}{5 c^5} \I^{ij(1)} \I_{ij}^{(5)}.
\end{align}
Evaluating the components of the reduced quadrupole moment together with their time derivatives yields
\begin{align} \label{AdEdt0a}
\frac{dE}{dt} = ~ & \frac{8 G^2 m^2}{15 c^5 r^4} \Big\{ 
G M(r)^2 \left( 8 \dot r^2 + 6 (r \dot \varphi)^2 \right) \\
& + r M(r) \Big( 6 \dot r^4 - 6 \left( 3 (r \dot \varphi)^4 + G (r \dot \varphi)^2 M' \right) \nonumber \\
& ~~~~ + \dot r^2 \left( 63 (r \dot \varphi)^2 + G M' - 3 G r M'' \right) \Big) \nonumber \\
& + r^2 \Big( 6 (r \dot \varphi)^4 M' \nonumber \\
& ~~~~ - \dot r^2 \left(G (M')^2 + 3 (r \dot \varphi)^2 \left(13 M' - 3 r M'' \right) \right) \nonumber \\ 
& ~~~~- \dot r^4 \left( 2 M' + r M'' - r^2 M''' \right) \Big) \Big\} \nonumber
\end{align}
(where we have omitted the argument $(r)$ for derivatives of the enclosed mass $M(r)$).  For a circular orbit in the stellar exterior, with $\dot r = 0$ and 
$\Omega \equiv \dot \varphi = (GM/r^3)^{1/2}$, this yields the familiar result 
$dE/dt = - 32 G^4 m^2 M^3/(5 c^5 r^5)$. 

%
\subsection{Gravitational-wave strains}
\label{sec:radreac:strain}
%

We next compute the gravitational-wave strains $h_+$ and $h_\times$ from the reduced quadrupole moments (\ref{AI}).  Focusing on an observer's location on the $z$-axis, so that the orbit of the black hole is seen face on, the two polarization amplitudes can be computed from
\begin{subequations}
\begin{align}
    h_+ = &~ \frac{G}{c^4 d} \left( \I_{xx}^{(2)} - \I_{yy}^{(2)} \right), \\
    h_\times = &~ \frac{2G}{c^4 d} \, \I_{xy}^{(2)},
\end{align}
\end{subequations}
where $d$ is the observer's distance from the source, and where the amplitudes and reduced quadrupole moments are measured at retarded time $t-d$.  Inserting the expressions (\ref{AI}), using the equations of motion (\ref{AEOM}) to replace second derivatives of $x$ and $y$, and finally expressing the result in terms of $r$ and $\varphi$ using $x = r \cos \varphi$ and $y = r \sin \varphi$ we obtain
\begin{subequations} \label{Ah1}
\begin{align}
    h_+ = &~ - \frac{2 G m}{c^4 d} \Big\{ \frac{G M(r) \cos(2 \varphi)}{r} - 
        \dot r^2 \cos(2 \varphi)  \\
        & ~~~~~~~  + 2 \dot r r \dot \varphi \sin(2 \varphi)  + (r \dot \varphi)^2 \cos(2 \varphi) \Big\} \nonumber \\
    h_\times = &~ - \frac{2 G m}{c^4 d} \Big\{ \frac{G M(r) \sin(2 \varphi)}{r}  \\
        & ~~~~~~~ - 2 \big( \dot r \sin\varphi + r \dot \varphi \cos\varphi \big) \big( \dot r \cos\varphi - r \dot \varphi \sin\varphi \big)    \Big\}. \nonumber 
\end{align}
\end{subequations}
For a circular orbit with $\dot r = 0$ and $\Omega = (G M(r)/r^3)^{1/2}$ this reduces to the more familiar expressions
\begin{subequations}
\begin{align}
    h_+ = &~ - \frac{4 G m r^2 \Omega^2 \cos(2 \varphi)}{c^4 d}  \\
    h_\times = &~ - \frac{4 G m r^2 \Omega^2 \sin(2 \varphi)}{c^4 d} 
\end{align}
\end{subequations}
with $\varphi = (t - d) \Omega$.

%
\subsection{Energy and angular momentum losses in the stellar exterior}
\label{sec:radreac:exterior}
%

The emission of energy and angular momentum over part of an orbit, e.g.~in the exterior, is purely gauge dependent and hence not a measurable physical quantity.  Only when averaged over many wavelengths, i.e.~over full orbits, are the emissions gauge-invariant and physically meaningful.  

In our simulations we compute orbital segments in the stellar interior numerically, where the losses in energy and angular momentum are accounted for by including the radiation-reaction forces (\ref{AFRR1}) in the equations of motion (\ref{eom}).  In order to obtain the losses from the entire orbit, we therefore have to add the contributions from the orbit in the stellar exterior.  Since we describe these exterior parts of the orbit analytically (see Section \ref{sec:numerics} and Appendix \ref{sec:exterior}), we also need to compute the energy and angular momentum losses from those parts analytically.  Combined with the contributions from the interior part of the orbit, they then provide physically meaningful expressions for the loss of energy and angular momentum.

In the exterior of the stellar host, the radiation reaction force (\ref{AFRR1}) simplifies because all derivatives of the enclosed mass $M(r)$ vanish.  Moreover, the point mass $m$ follows, to leading order, a Keplerian orbit.  Once it is captured gravitationally, its orbit in the stellar exterior is a segment of an ellipse with semi-major axis $a$ and eccentricity $e < 1$.  The binary separation $r$ is given by
\begin{equation} \label{Ar}
r = \frac{a(1 - e^2)}{1 + e \cos \varphi},
\end{equation}
where $\varphi = 0$ corresponds to the pericenter, and the angular velocity can be written as 
\begin{equation} \label{Aphidot}
    \dot \varphi = \frac{L}{m r^2} 
    = \frac{\left(G M a (1 - e^2) \right)^{1/2}}{r^2}
\end{equation}
(where we have have again assumed $m \ll M$).  Using these expressions we can now find the emitted energy and angular momentum analytically (compare \cite{PetM63,Pet64}). 

\subsubsection{Emission of energy}

The rate at which the point-mass $m$ loses energy can be computed from (\ref{AdEdt2a}).  Inserting the radiation-reaction force (\ref{AFRR1}) together with (\ref{Ar}) and (\ref{Aphidot}) yields
\begin{align} \label{AdEdt1a}
\frac{dE}{dt} = & - \frac{G^4}{15c^5}  \frac{m^2 M^3}{a^5 (1 - e^2)^5} (1 + e \cos \varphi)^3 \nonumber \\
& ~~~ \times \Big(96 + 76 e^2 - 27 e^4 + 16 e (27 + 8 e^2) \cos(\varphi)
\nonumber \\
& ~~~~~~~~    + 4 e^2 (161 + 24 e^2) \cos (2 \varphi) + 400 e^3 \cos(3 \varphi)
\nonumber \\ 
& ~~~~~~~~ + 75 e^4 \cos(4 \varphi) \Big),
\end{align}
which, for a circular orbit with $e = 0$, again reduces to the well-known result $dE / dt = - 32 G^4 m^2 M^3 / (5 c^5 a^5)$.

As an aside, we note that (\ref{AdEdt1a}) is different from the corresponding equation (15) in \cite{PetM63} (hereafter PM15).  There are two reasons for these differences.  One of these reasons is quite simple -- we adopt $m \ll M$ in the equations of motion (\ref{AEOM}) and in (\ref{Aphidot}), while \cite{PetM63} do not make this assumption -- but the second reason is more subtle.   Specifically, rather than computing the loss of energy from the forces $F^i_{RR}$ in (\ref{AdEdt2a}), as we did above, we could find it from the quadrupole moment $\I_{ij}$ and its first and fifth derivatives -- as we did in Section \ref{sec:radreac:general} in order to obtain (\ref{AdEdt0a}).  In calculations aimed at computing the {\em time-averaged} emission of energy over one or many orbits -- as, for example, in \cite{PetM63,Pet64} -- it is possible to integrate these expressions by parts twice and ignore boundary terms.  This brings the last term in (\ref{AdEdt2a}) into the more familiar form
\begin{equation} \label{AdEdt2b}
\Bigl \langle \frac{dE}{dt} \Bigr \rangle = - \frac{G}{5 c^5} \I^{ij(3)} \I_{ij}^{(3)},
\end{equation}
which, requiring only third rather than fifth time derivatives, is easier to evaluate.  The expression (PM15) is based on (\ref{AdEdt2b}), and hence assumes an averaging that results in the disappearance of boundary terms, whereas our expression (\ref{AdEdt1a}) does not.  We have verified (a) that we obtain (PM15) from (\ref{AdEdt2b}) and (\ref{AI}), and (b) that starting with the last line in (\ref{AdEdt2a}) (and assuming $m \ll M$) also results in (\ref{AdEdt1a}).

The energy $\Delta E$ lost between two angles $\varphi_1$ and $\varphi_2$ can then be found by integrating
\begin{equation} \label{AEint}
\Delta E = \int \frac{dE}{dt} dt = \int \frac{dE / dt}{d \varphi / dt} d \varphi = \frac{m}{L} \int_{\varphi_1}^{\varphi_2} r^2 \frac{dE}{dt} d \varphi,
\end{equation}
where we have used (\ref{Aphidot}) in the last equality and approximated $L$ as constant to leading order.  

In order to find the energy emitted in the exterior of the star we integrate from the angle $\varphi_1 = \varphi_{\rm surf}$ corresponding to the surface $r = r_*$, which can be found from (\ref{Ar}), to the apcenter at $\varphi_2 = \pi$, and multiply by two to find
\begin{subequations} \label{ADeltaE}
\begin{align}
\Delta E = & - \frac{G^{7/2}}{90 c^5}
\frac{m^2 M^{5/2}}{a^{7/2}} g_{\rm E}(e, \varphi_{\rm surf}),
\end{align}
where we have defined the dimensionless function
\begin{align}
& g_{\rm E}(e, \varphi_{\rm surf}) \equiv 
\Big\{ 
    12 \left(96 + 292 e^2 + 37 e^4 \right) (\pi -\varphi_{\rm surf} ) \nonumber \\
   & ~~~~~ - 12 e (24 + e^2)(22 + 21 e^2) \sin (\varphi_{\rm surf}) \nonumber \\
   & ~~~~~ - 120 e^2 (43 + 18 e^2) \sin(2 \varphi_{\rm surf}) 
   \nonumber \\
   & ~~~~~~ - 38 e^3 (76 + 9 e^2 ) \sin(3 \varphi_{\rm surf})
   - 825 e^4 \sin(4 \varphi_{\rm surf}) \nonumber \\
   & ~~~~~~ - 90 e^5 \sin(5 \varphi_{\rm surf}) 
   \Big\} / (1 - e^2)^{7/2}.
\end{align}
\end{subequations}

As a consistency check we can recover the result for a complete orbit by inserting $\varphi_{\rm surf} = 0$.  For this special case we may then divide by the orbital period
\begin{equation} \label{Aperiod}
T = \frac{2 \pi}{G^{1/2} M^{1/2}} a^{3/2}
\end{equation}
to obtain the loss of energy averaged over one orbit
\begin{equation} \label{AdEdt_ave}
\Bigl \langle \frac{dE}{dt} \Bigr \rangle = - \frac{32 G^4}{5 c^5}
\frac{m^2 M^3}{a^5 (1 - e^2)^{7/2}} 
\left(1 + \frac{73}{24} e^2 + \frac{37}{96} e^4 \right). 
\end{equation}
This expression {\em does} agree with the corresponding expression (PM16) (as well as 5.4 in \cite{Pet64}, hereafter P5.4), 
evaluated in the limit $m \ll M$.  This was to be expected, because now the integration {\em is} taken over an entire orbit, so that the boundary terms neglected in (\ref{AdEdt2b}) do indeed vanish, making it equivalent to our starting point (\ref{AdEdt2a}).  The result is a gauge-invariant and physically measurable quantity (at least in principle).

\subsubsection{Emission of angular momentum}

We now adopt a similar approach to compute the loss in angular momentum (about the $z$-axis) from
\begin{equation}
    \frac{dL}{dt} = ({\bf r} \times {\bf F}_{\rm RR})^z = r^2 F_{\rm RR}^\varphi,
\end{equation}
Inserting (\ref{AFRRphi}) together with (\ref{Ar}) and (\ref{Aphidot}) we find
\begin{align}
\frac{dL}{dt} = &  -\frac{4 G^{7/2}}{5 c^5} \frac{m^2 M^{5/2}}{a^{7/2} (1 - e^2)^{7/2}} \left(1 + e \cos(\varphi)\right)^3 \nonumber \\
& ~~~ \times (8 - 3 e^2 + 20 e \cos (\varphi) + 15 e^2 \cos(2 \varphi) )
\end{align}
In analogy to (\ref{AEint}) we now integrate
\begin{equation}
    \Delta L = \frac{2 m}{L} \int_{\varphi_{\rm surf}}^{\pi} r^2 \frac{dL}{dt} d \varphi
\end{equation}
to find the angular momentum radiated in the exterior of the star,
\begin{subequations} \label{ADeltaL}
\begin{align} 
    \Delta L = & - \frac{8 G^3}{5 c^5} \frac{m^2 M^2}{a^2} 
    g_{\rm L}(e, \varphi_{\rm surf})
\end{align}
with
\begin{align}
& g_{\rm L}(e, \varphi_{\rm surf})
\equiv \frac{1}{(1 - e^2)^2}
\Big\{ (8 + 7 e^2)(\pi - \varphi_{\rm surf}) \nonumber \\
    & ~~~~~ - e \sin(\varphi_{\rm surf}) \Big(28 + 7 e^2 + 25 e \cos(\varphi_{\rm surf}) \nonumber \\
    & ~~~~~ + 5 e^2 \cos(2 \varphi_{\rm surf}) \Big) \Big\}
\end{align}
\end{subequations}
As for the emitted energy we can check our result by computing the energy emitted in a complete orbit (using $\varphi_{\rm surf} = 0$) and dividing by the period (\ref{Aperiod}) to obtain the orbital average
\begin{equation}
\Bigl \langle \frac{dL}{dt} \Bigr \rangle = - \frac{32 G^{7/2}}{5 c^5} 
\frac{m^2 M^{5/2}}{a^{7/2} (1 - e^2)^2} \left(1 + \frac{7}{8} e^2 \right).
\end{equation}
This agrees with (P5.5) as expected.

As we stated above, we add the changes (\ref{ADeltaE}) and (\ref{ADeltaL}) in the energy and angular momentum to those obtained from the numerical integration of the equations of motion (\ref{eom}) in order to obtain physically meaningful expressions.

%
\section{Estimates for the Mach number}
\label{sec:mach}
%

In this appendix we provide some analytical, Newtonian estimates for the Mach number $\mach$ of a PBH inside a host neutron star.

\subsection{Initial transit}

Neglecting dissipative forces and assuming $v_\infty \ll v_{\rm esc}$, so that we may approximate the PBH's energy per unit mass $E$ to vanish, we can find its speed $v$ inside the neutron star from
\begin{equation} \label{C_E}
E = \frac{1}{2} v^2 + \Phi(r) = 0,
\end{equation}
where $\Phi(r)$ is the Newtonian potential.  

Focusing on a $\Gamma = 2$ polytrope, the enclosed mass $M(r)$ in the stellar interior is given by
\begin{equation} \label{C_mass}
     M(r) = M_* 
     \frac{\sin \xi - \xi \cos \xi}{\pi}
\end{equation}
where we have defined 
\begin{equation} \label{C_xi}
\xi \equiv \frac{r}{r_*} \pi.  
\end{equation}
From (\ref{C_mass}) we can compute the potential
\begin{equation}
\Phi(r) = - \frac{M_*}{r_*} \left( 1 + \frac{\sin \xi}{\xi} \right),
\end{equation}
which we insert into (\ref{C_E}) to obtain
\begin{equation}
v^2 = \frac{2 M_*}{r_*} \left( 1 + \frac{\sin \xi}{\xi} \right).
\end{equation}
We note that the speed at the stellar center, where $\xi = 0$, is larger than that at the surface by a rather modest factor of $v(0)/v(r_*) = \sqrt{2}$.

In the Newtonian limit, the sound  speed (\ref{a_of_rho}) reduces to
\begin{equation} \label{C_a}
a = \left(\Gamma K \rho^{\Gamma - 1}\right)^{1/2},
\end{equation}
where we no longer distinguish between the rest-mass density $\rho_0$ and the total energy density $\rho$.  We now specialize to $\Gamma = 2$ again, use $\rho / \rho_c = \sin \xi / \xi$ as well as $M/R = 2 K \rho_c$ for Newtonian $\Gamma = 2$ polytropes, where $\rho_c$ is the central density, and obtain the Mach number
\begin{equation}
\mach = \frac{v}{a} = 2^{1/2} \left(1 + \frac{\xi}{\sin \xi} \right)^{1/2}. 
\end{equation}
As expected, we find that $\mach = \infty$ at the stellar surface where $\xi = \pi$, since the density and hence the sound speed vanish there.  The Mach number decreases as the PBH falls into the star, and takes its minimum value of 
\begin{equation}
\mach_{\rm min} = 2
\end{equation}
at the stellar center (see also the discussion and Fig.~1 in \cite{GenST20}).  The initial transit of orbits with $v_{\infty} \ll v_{\rm esc}$ is therefore entirely supersonic during its passage through the neutron star.

\subsection{Circular orbits}

Even though we do not expect non-circular orbits to circularize as the PBH sinks to the stellar center, so that we expect nearly circular orbits to arise only from near-grazing capture, we provide here a simple estimate for the Mach number of PBHs in circular orbits inside host stars.  

Using (\ref{C_mass}) and (\ref{C_xi}) we first compute the orbital speed from
\begin{equation} \label{C_v_circ}
v = \left( \frac{M(r)}{r} \right)^{1/2}
= \left( \frac{M_*}{r_*} \frac{\sin \xi - \xi \cos \xi}{\xi} \right)^{1/2}.
\end{equation}
We then combine (\ref{C_v_circ}) with (\ref{C_a}) and use the same identities for Newtonian $\Gamma = 2$ polytropes as above to obtain
\begin{equation}
\mach = \left( \frac{\xi}{\sin \xi} \frac{\sin \xi - \xi \cos \xi}{\xi} \right)^{1/2} = \left(1 - \xi \cot \xi \right)^{1/2}.
\end{equation}
As expected, we find $\mach \rightarrow \infty$ toward the stellar surface, and $\mach \rightarrow 0$ for circular orbits close to the stellar center.  In particular, this indicates that for such orbits the PBH will move subsonically eventually.

%
\section{Approximate relation between $m$ and $a$}
\label{sec:m_versus_a}
%

In this brief appendix we provide an approximate Newtonian argument to motivate the power-law relation between $m$ and $a$ observed in Fig.~\ref{fig:m_versus_a} and the fit (\ref{m_versus_a_fit}).

We assume that the dissipative forces are dominated by accretion (see Fig.~\ref{fig:finalforces}) in which case we may write the Newtonian equation of motion as
\begin{equation}
m \ddot{\bf r} = - \frac{G M(r)}{r^2} \hat{\bf r} - \dot m \dot{\bf r}.
\end{equation}
Taking the dot product with the velocity $\dot {\bf r}$ we rewrite this as
\begin{equation}  \label{D1}
\frac{1}{2} m \frac{d v^2}{dt} + \frac{G M(r)}{r^2} \dot r = - \dot m v^2, 
\end{equation}
where we have used $v^2 = \dot{\bf r} \cdot \dot{\bf r}$.

We next specialize to a circular orbit with semi-major axis $a = r$ and speed
$v^2 = G M(a)/a$.  Focusing on the vicinity of the stellar center, where we may approximate the density $\rho$ as constant and hence the enclosed mass $M(r)$ as $M(r) \simeq 4 \pi \rho r^3 / 3$, the speed is given by
\begin{equation} \label{D2}
v^2 \simeq \kappa a^2,
\end{equation}
where we have defined $\kappa \equiv 4 \pi G \rho / 3$ for convenience.  Using (\ref{D2}) and the definition of $\kappa$ in (\ref{D1}) then yields
\begin{equation}
\frac{1}{2} \kappa m \frac{d a^2}{dt} + m \kappa a \dot a \simeq - \dot m \kappa a^2,
\end{equation}
which we may rewrite as 
\begin{equation}
\frac{\dot m}{m} \simeq - 2 \frac{\dot a}{a}
\end{equation}
and integrate to obtain
\begin{equation}
m \propto a^{n}
\end{equation}
with $n = -2$.

We have verified numerically that we obtain this power-law scaling for weak-field configurations sufficiently close to the stellar center when dynamical friction is turned off and dissipation is dominated by accretion.  Including dynamical friction and allowing for effects of strong-field gravity results in departures of $n$ from the value of -2 found here, as observed in Fig.~\ref{fig:m_versus_a}.

\end{appendix}


%

\end{document}